\documentclass[aps,pra,superscriptaddress,twocolumn]{revtex4-2}
\usepackage{amsmath}
\usepackage{amssymb}
\usepackage{graphicx}
\usepackage{epsfig}
\usepackage{adjustbox}
\usepackage{amstext}
\usepackage{esint}
\usepackage{color}
\usepackage{soul}
\usepackage{diagbox}
\usepackage[colorlinks=true, linkcolor=blue, citecolor=blue, urlcolor=blue]{hyperref}
\usepackage{tikz}
\usetikzlibrary{quantikz}

\begin{document}

\title{Digital Quantum Simulation of the Holstein-Primakoff Transformation on Noisy Qubits}

\author{Kelvin Yip}
\affiliation{School of Natural Sciences, University of California, Merced, California 95343, USA}

\author{Alessandro Monteros}
\affiliation{School of Natural Sciences, University of California, Merced, California 95343, USA}

\author{Sahel Ashhab}
\affiliation{Advanced ICT Research Institute, National Institute of Information and Communications Technology (NICT), 4-2-1 Nukui-Kitamachi, Koganei, Tokyo 184-8795, Japan}
\affiliation{Research Institute for Science and Technology, Tokyo University of Science, 1-3 Kagurazaka, Shinjuku-ku, Tokyo 162-8601, Japan}
\affiliation{Department of Physics, The University of Tokyo, Tokyo 113-0033, Japan}

\author{Lin Tian}
\email{ltian@ucmerced.edu}
\affiliation{School of Natural Sciences, University of California, Merced, California 95343, USA}

\begin{abstract}
Quantum simulation of many-body systems offers a powerful approach to exploring collective quantum dynamics beyond classical computational reach. Although spin and fermionic models have been extensively simulated on digital quantum computers, the simulation of bosonic systems on programmable quantum processors is often hindered by the intrinsically large Hilbert space of bosonic modes. In this work, we study the digital quantum simulation of bosonic modes using the Holstein-Primakoff (HP) transformation and implement this protocol on a cloud-based superconducting quantum processor. Representative bosonic dynamics, including the driven harmonic oscillator, bosonic squeezing, and the Jaynes-Cummings (JC) model, are simulated on quantum hardware. Using data obtained from the quantum simulations, we systematically examine the interplay between algorithmic and hardware-induced errors to identify optimal simulation parameters. The dominant algorithmic errors arise from the finite number of qubits used in the HP mapping and the finite number of Trotter steps in the time evolution, while hardware errors mainly originate from entangling-gate infidelity, decoherence, and readout errors. We demonstrate two-excitation JC dynamics and squeezed-state generation involving qubit-qubit interactions under the HP transformation, extending the simulations beyond the simple single-excitation regime. Furthermore, we show that quantum error mitigation substantially improves the fidelity of the quantum simulations on current hardware. Our results demonstrate the feasibility and limitations of HP-based digital simulation of bosonic degrees of freedom on current quantum processors and provide a framework that can be extended to more complex spin-boson and multimode cavity models.
\end{abstract}
\maketitle

\section{Introduction \label{sec:intro}}
Digital and analog quantum simulations of many-body systems on programmable quantum processors provide a versatile route for exploring complex quantum dynamics that are intractable for classical computation~\cite{GeorgescuRMP2014QS, TacchinoAQT2020, Daley2022practical, FausewehNatCommon2024ReviewDigitalQS}. 
In recent years, cloud-based quantum platforms such as IBM Quantum (IBMQ)~\cite{IBMQC} have enabled rapid advances in the demonstration of digital quantum simulation~\cite{KandalaNature2017, FreyRachelScienceAdv2022TimeCrystalIBMQ, KimNature2023, YoshiokaNatCommon2025, OftelieQST2021, AbhijithACMTransQC2022ImplementationQAlgorithms, AbdurakhimovPreprint2024IQM}. 
Leveraging Qiskit and related open-source toolkits~\cite{IBMQiskit}, a wide variety of models including spin chains and molecular Hamiltonians have been implemented on cloud quantum processors~\cite{LammPRL2018NonequilibriumDynamics, ZhukovQIP2018algorithmic, ChiesaNatPhys2019, SmithNpjQI2019, FausewehQIP2021NonEquilibrium, ChowdhuryPRR2024LargeScaleIBMQ}. 
The accessibility of cloud-based quantum processors enables systematic benchmarking of simulation algorithms, circuit optimization strategies, noise characterization and modeling, and error-mitigation techniques on real quantum hardware, whose performance is limited by finite qubit coherence times, non-negligible entangling-gate errors, and readout errors~\cite{GeorgopoulosPRA2021,UrbanekPRL2021, DiBartolomeoPRR2023, DahlhauserPRA2021, Bravo-MontesEPJQT2024, BordoniQST2025NoiseModellingReinforcedLearning, MagesanPRL2011, DesdentadoComputing2025QuantumSoftwareHardware, Nguyen2024arXivQuantumCloudComputingReview, TemmePRL2017, Giurgica-TironIEEE2020, AHePRA2020ErrorMitigation, NationPRXQuantum2021, RosenbergQST2022ErrorMitigation, ZCaiRMP2023ErrorMitigationReview}.

Quantum simulation of bosonic models using digital quantum processors is fundamentally challenging due to the intrinsic structure of their Hilbert space.  In contrast to spin and fermionic systems, bosonic modes possess an infinitely large Hilbert space~\cite{Nielsen_Chuang_2010}. As a result, it is difficult to implement digital quantum simulations of such large Hilbert spaces on noisy quantum processors, which are limited to finite circuit depth and coherence times. To overcome this challenge, analog and hybrid quantum simulation approaches have been employed to mimic models involving bosonic modes by exploiting other bosonic systems, such as superconducting resonator modes and the harmonic motion of trapped ions~\cite{StavengerPreprint2209.11153BostonQiskit, MezzacapoSciRep2014, BabukhinPRA2020}. These approaches provide a natural solution for simulating bosonic models and have been applied to a broad range of systems, spanning models in cavity and circuit quantum electrodynamics setups, realizations of Bose-Fermi mixture models, and strongly correlated bosonic many-body Hamiltonians~\cite{LangfordNatCommun2017, ShapiroPRA2025, BurgerEntropy2022, FMeiPRB2013, SeoPRB2015, KCaiNpjQI2021, KCaiPRA2024}. Meanwhile, digital quantum simulation has also been employed to emulate bosonic systems with a bounded number of excitations and has been implemented across various physical platforms, including nuclear magnetic resonance systems, superconducting qubits, and trapped ions~\cite{SomarooCoryPRL1999, SawayaNpjQI2020, SabinQuantumReport2020, EncinarPRA2021, SommaThesis_arXiv0512209, BaishyaPreprint1906.01436QHODriven, SChinPreprint2209.00207SimulatingBosons, XYHuang2105.12563SimulatingBosons, MarinkovicSciRep2023MappingNoisyQubits, RubinUCD2404.03861QST2025}. In these digital approaches, the bosonic Hilbert space is truncated to a finite (and often small) number of excitations, with the Fock states of the bosonic modes mapped onto multi-qubit states. Another theoretical framework that bridges bosonic modes and spin ensembles is the Holstein-Primakoff (HP) transformation~\cite{HolsteinPhysRev1940, ArecciPRA1972}, which maps bosonic operators to collective angular momentum operators of spin ensembles, or vice versa.  Through this mapping, a bosonic mode can be simulated by a qubit ensemble, enabling the qubits to reproduce bosonic dynamics and emulate a wide range of boson-related phenomena. The HP transformation provides a systematic and physically motivated truncation of the bosonic Hilbert space. Compared with a binary encoding, which requires only $\sim\log_2(N+1)$ qubits to represent $N+1$ bosonic states, the HP mapping is more costly in qubit resources and requires $N$ qubits. In return, it yields simple operators: linear bosonic terms map onto single-qubit rotations, and couplings to other spins become two-qubit interactions, whereas in a binary encoding the bosonic operators become multi-qubit Pauli strings whose implementation generally requires additional entangling gates. The HP mapping can therefore be advantageous when circuit depth, rather than the number of qubits, is the main limitation and the excitation number remains small compared with $N$. Recent work has employed this approach to simulate spin-boson models, in which both the bosonic (cavity) mode and the spins are encoded using qubits~\cite{TudorovskayaPRA2024, FitzpatrickPreprint2106.03985FermionBosonSystem}. 

In this work, we investigate the digital quantum simulation of the HP transformation using noisy qubits on a cloud-based superconducting quantum processor (IBMQ). By mapping a bosonic mode onto an ensemble of superconducting qubits, we construct quantum circuits that reproduce the time evolution of the bosonic mode. We simulate representative bosonic dynamics, including the driven harmonic oscillator and bosonic squeezing, as well as the Jaynes-Cummings (JC) model describing light-matter interaction. We investigate two-excitation JC dynamics and squeezed-state generation involving qubit-qubit interactions under the HP transformation, thereby demonstrating the applicability of the HP encoding beyond simple single-excitation dynamics. Using these simulations, we systematically examine the effects of algorithmic errors arising from the finite number of qubits used in the HP transformation and the finite number of Trotter steps employed in the time evolution, together with hardware noise sources such as finite entangling-gate fidelities and readout errors. By comparing noiseless classical simulations performed with Qiskit with results obtained on real quantum hardware, including results with quantum error mitigation (QEM), we identify optimal parameter regimes that maximize the simulation fidelity. In addition, we employ a synthesized unitary approach to construct optimized quantum circuits for implementing the JC model. Our results demonstrate the feasibility and limitations of HP-based digital simulation of bosonic modes on current noisy quantum hardware and provide a framework for extending these simulations to more complex spin-boson and multimode systems.

The remainder of this paper is organized as follows. In Sec.~\ref{sec:HPT}, we introduce the HP transformation, which maps a bosonic mode onto a spin ensemble, and describe how bosonic operators are represented by collective qubit operators. In Sec.~\ref{sec:dho}, we first perform a quantum simulation of the driven harmonic oscillator using this mapping and analyze the effects of the finite number of qubits used in the HP transformation as well as the impact of readout noise on the simulation. We further present results on the simulation of bosonic squeezing using the HP transformation, providing a more involved test of the HP-based simulation that requires qubit-qubit interactions. In Sec.~\ref{sec:JCM}, we simulate the quantum dynamics of the JC model with one and two excitations by employing both the Suzuki-Trotter decomposition~\cite{Suzuki1990PhysLettA} and the synthesized unitary approach~\cite{Goubault2019synthesizing, Ashhab2022numerical, Ashhab2024quantum, BablerQuantum2023}. The impact of entangling-gate error on the performance of these simulations is also assessed. Throughout Secs.~\ref{sec:dho}--\ref{sec:JCM}, we apply QEM, including zero-noise extrapolation for entangling-gate errors and measurement error correction, to improve the hardware-simulated results. Finally, Sec.~\ref{sec:conclusion} presents the conclusion of this work.

\section{HP Transformation \label{sec:HPT}}
Consider an ensemble of $N$ spins described by the Pauli spin operators $\sigma_{i\pm}$ and $\sigma_{iz}$, where $i \in [1,\,N]$. The collective spin operators are defined as ($\hbar = 1$)
\begin{equation}
    J_z = \frac{1}{2}\sum_{i=1}^N \sigma_{iz}, 
    \qquad 
    J_{\pm} = \sum_{i=1}^N \sigma_{i\pm},
    \label{eq:J_def}
\end{equation}
which are sums of individual spin operators and satisfy the standard angular momentum commutation relations 
$[J_{+},\,J_{-}] = 2J_z$ and $[J_{z},\,J_{\pm}] = \pm J_{\pm}$. In the HP transformation~\cite{HolsteinPhysRev1940}, the collective spin operators are mapped onto the operators of a single bosonic mode, or vice versa, according to
\begin{subequations}\label{eq:HPT1}
    \begin{align}
        J_{+} &= a^{\dagger}\sqrt{N - a^{\dagger}a}, \\
        J_{-} &= \sqrt{N - a^{\dagger}a}\, a, \\
        J_z &= a^{\dagger}a - \frac{N}{2},
    \end{align}
\end{subequations} 
where $a$ ($a^{\dagger}$) is the annihilation (creation) operator of the bosonic mode, satisfying the commutation relation $[a,\,a^{\dagger}] = 1$. Under this mapping, the bosonic ground state $\ket{0}$ corresponds to the fully polarized spin-down state $ \ket{J=\tfrac{N}{2},\,J_z=-\tfrac{N}{2}} = \ket{\downarrow\downarrow\cdots\downarrow}$. The $n$th Fock state corresponds to the Dicke state $\ket{D_n^N}$ defined as the fully permutation-symmetric superposition of all spin configurations with $n$ spins pointing up and the remaining $N-n$ spins pointing down. These Dicke states form an effective bosonic ladder under the HP mapping, where each excitation corresponds to a single collective spin flip within the ensemble. For noiseless, ideal qubits, the bosonic dynamics are restricted to the Dicke sector with $J=N/2$, which is preserved by the symmetric Hamiltonian evolution implemented in the quantum circuits. For noisy, realistic systems, however, hardware errors can break the symmetry and induce leakage into nonsymmetric states.

\begin{figure}
    \centering
    \includegraphics[width=\columnwidth, clip]{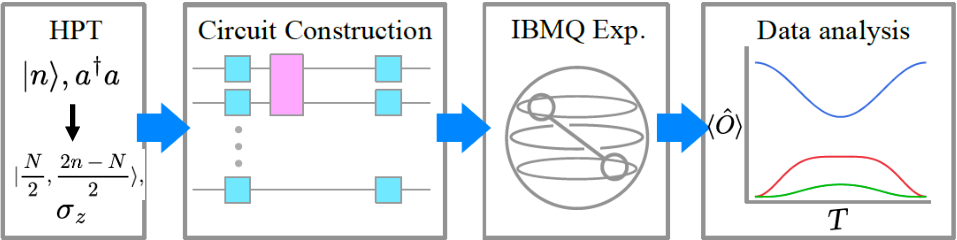}
    \caption{Schematic workflow of the quantum simulation procedure, including the mapping of a bosonic mode onto a qubit ensemble, quantum circuit construction, IBMQ execution, and analysis of the measured data (including QEM). Both noiseless classical simulations and quantum hardware experiments are performed.} 
    \label{fig1}
\end{figure}

In the large $N$ limit with $N \gg \langle a^{\dagger}a \rangle$, the square-root factors in Eq.~\eqref{eq:HPT1} can be approximated by $\sqrt{N}$, yielding
\begin{equation}
    a^{\dagger} \approx \frac{J_{+}}{\sqrt{N}},
    \qquad
    a \approx \frac{J_{-}}{\sqrt{N}}.
    \label{eq:HPT2}
\end{equation}
In this regime, the bosonic creation and annihilation operators are effectively represented by normalized collective spin-raising and spin-lowering operators. The accuracy of this approximation improves as $N/\langle a^{\dagger}a\rangle$ increases and becomes exact in the limit $N\rightarrow\infty$ for a finite excitation number. Using this correspondence, we investigate the quantum simulation of single-mode bosonic dynamics, including the driven quantum harmonic oscillator and bosonic squeezing, as well as the spin-boson dynamics of the JC model, and implement these simulations on IBMQ~\cite{IBMQC}. The overall workflow of the simulations is illustrated in Fig.~\ref{fig1}.

\section{Digital Simulation of Single-Mode Bosonic Systems \label{sec:dho}} 

\subsection{Bosonic Models and HP Transformation}
First, we consider a driven quantum harmonic oscillator described by the Hamiltonian
\begin{equation}
    H = \omega_0 \left(a^{\dagger}a + \tfrac{1}{2}\right)
      + F \cos(\omega_d t)\,(a + a^{\dagger}),
    \label{eq:Hdho}
\end{equation}
where $\omega_0$ is the oscillator frequency, $\omega_d$ is the driving frequency, and $F$ is the driving amplitude. Assuming $\omega_0,\,\omega_d \gg |\omega_0 - \omega_d|,\, F$, the rotating-wave approximation (RWA) can be applied. In the rotating frame defined by $H_0 = \omega_d a^{\dagger} a$, the Hamiltonian under the RWA becomes
\begin{equation}
    H_{\mathrm{rot}} = \Delta\, a^{\dagger}a + \tfrac{F}{2}(a + a^{\dagger}),
    \label{eq:Hdho_rot}
\end{equation}
where $\Delta = \omega_0 - \omega_d$ is the detuning between the oscillator and the drive, and a constant term $\tfrac{\omega_0}{2}$ has been omitted. The driving term effectively shifts the origin of the harmonic potential by 
$d_0 = -F/(\sqrt{2}\,\Delta)$ along the coordinate axis. Here the coordinate operator of the harmonic oscillator has the form $x = \tfrac{1}{\sqrt{2}}(a + a^{\dagger})$ in terms of the creation and annihilation operators. Starting from the vacuum state $\ket{0}$, the oscillator evolves into a coherent state $\ket{\alpha(t)}$ that oscillates periodically about the displaced origin, with
\begin{equation}
    \alpha(t) = \frac{d_0}{\sqrt{2}}\!\left[1 - \cos(\Delta t) + i\sin(\Delta t)\right].
\end{equation}
The probability of occupying the $n$th Fock state of the oscillator mode is
\begin{equation}
    P_{\mathrm{b}}^{(n)}  = |\langle n | \alpha(t) \rangle|^2 
         = e^{-|\alpha|^2} \frac{|\alpha|^{2n}}{n!},
    \label{eq:Pn}
\end{equation}
where $|\alpha(t)|^2 = 2d_0^2 \sin^2(\tfrac{\Delta t}{2})$.

When mapped onto qubit operators using the HP transformation in the large-$N$ limit, Hamiltonian~\eqref{eq:Hdho_rot} becomes
\begin{equation}
    H_{\mathrm{rot}} 
    = \sum_{i=1}^{N} \left[
        \tfrac{\Delta}{2}\,\sigma_{iz}
        + \tfrac{F}{2\sqrt{N}}\,\sigma_{ix}
      \right]
    = \sum_{i=1}^{N} \Omega\,\sigma_{i\hat{n}},
    \label{eq:Hdq_rot}
\end{equation}
where $\Omega = \sqrt{(\tfrac{\Delta}{2})^2 + (\tfrac{F}{2\sqrt{N}})^2}$, $\sigma_{i\hat{n}}$ denotes the Pauli operator along the rotated axis $\hat{n}$, and the constant term $\tfrac{N\Delta}{2}$ has been neglected. This Hamiltonian contains no qubit-qubit interactions and can therefore be implemented using single-qubit quantum logic gates only. For a single qubit initially prepared in the spin-down state, the time-dependent wavefunction under Eq.~\eqref{eq:Hdq_rot} is $\ket{\psi_i(t)} = e^{-i \Omega \sigma_{i\hat{n}} t}\ket{\downarrow}_i$. The probability of finding the qubit in the spin-up state is
\begin{equation}
    P_{\uparrow} = \frac{F^2}{4N\Omega^2}\sin^2(\Omega t).
    \label{eq:PupDQHO}
\end{equation}
When the Hamiltonian acts on $N$ qubits all initialized in $\ket{\downarrow}$, the probability of measuring $n$ spins in the state $\ket{\uparrow}$, which corresponds to the oscillator Fock state $\ket{n}$, is given by the binomial distribution
\begin{equation}
    P_{\mathrm{c}}^{(n)} 
    = \binom{N}{n} P_{\uparrow}^{\,n} (1 - P_{\uparrow})^{N-n}.
    \label{eq:PupN}
\end{equation}
It can be shown that in the limit of $N \gg n$, $P_{\mathrm{c}}^{(n)}$ approaches $P_{\mathrm{b}}^{(n)}$ in Eq.~\eqref{eq:Pn} when $|\alpha|^2 = N P_{\uparrow}$. In fact, for $|F/\Delta| \ll 1$ (ensuring a small excitation number), we have $|\alpha|^2 \approx N P_{\uparrow}$. Therefore, the excitation statistics of the spin ensemble reproduce the state distribution of the oscillator under the HP mapping. The implementation of this model is presented in Sec.~\ref{ssec:dho_implement}.

We next extend our study to the generation of squeezed states, which constitute an important class of nonclassical states of bosonic modes. Unlike coherent states, the generation of squeezed states involves the bilinear operators $a^2$ and $(a^\dagger)^2$, which are mapped to qubit-qubit interactions under the HP transformation. Consequently, implementing bosonic squeezing requires two-qubit gates and provides a more involved test of the digital quantum simulation of the HP transformation. The squeezing operator is defined as
\begin{equation}
S(r,\phi)=\exp\!\left[
\frac{r}{2}
\left(
e^{-2i\phi}a^{2} - e^{2i\phi}a^{\dagger 2}
\right)
\right],
\end{equation}
where $r$ and $\phi$ denote the squeezing amplitude and phase, respectively. For a squeezed state $S(r,\phi)\ket{0}$, the mean excitation number is $n_{\rm ex}^{({\rm b})} = \sinh^{2}r$.

Using the large-$N$ approximation for the HP transformation in Eq.~\eqref{eq:HPT2}, the squeezing operator is mapped onto qubit operators as
\begin{equation}
S(r,\phi)
\approx
\exp\!\left[
\frac{r}{2N}
\sum_{i\ne j}
\left(
e^{-2i\phi}\sigma_{i-}\sigma_{j-} - e^{2i\phi}\sigma_{i+}\sigma_{j+}
\right)
\right],
\label{eq:squeezing0}
\end{equation}
where $N$ is the number of qubits. The resulting operator contains noncommuting qubit-qubit coupling terms. Its digital implementation is discussed in Sec.~\ref{ssec:squeezing_implement}.

\begin{figure}
    \centering
    \includegraphics[width=\columnwidth, clip]{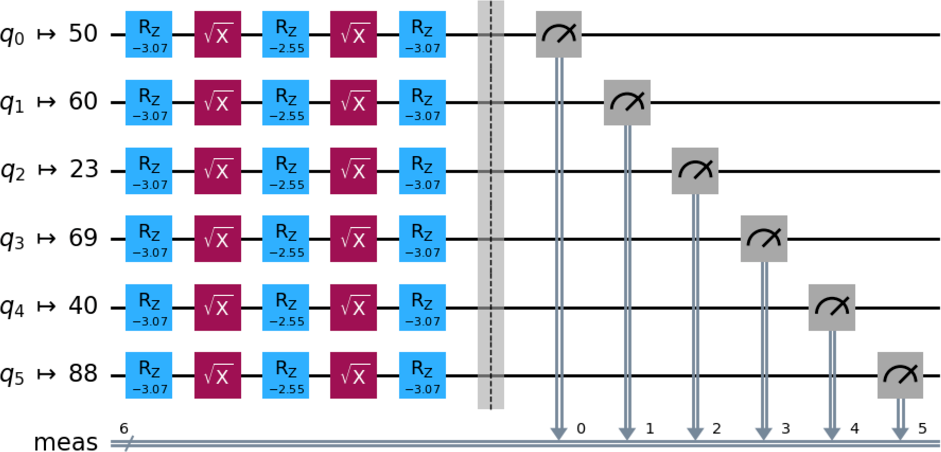}
    \caption{Circuit diagram for the digital simulation of the driven harmonic oscillator using six qubits, up to a global phase of $\pi$. The qubit labels (e.g., 50 and 60) indicate their physical indices on IBM Kingston~\cite{IBMQC}. The circuit is transpiled into the native IBMQ gate set and followed by qubit measurements. The rotation angles of the $R_z$ gates correspond to the parameters $\Delta = 1$, $F = 0.75$, and evolution time $t = \pi$, and are shown inside the blue boxes of the gates. All parameters are expressed in dimensionless units.}
    \label{fig2}
\end{figure}
\subsection{Circuit Implementation of Driven Harmonic Oscillator \label{ssec:dho_implement}}
We perform classical simulations of the HP mapping for the driven harmonic oscillator using Qiskit Aer and execute quantum simulations on IBM Kingston for a range of qubit numbers. Since the Hamiltonian of the driven harmonic oscillator [Eq.~\eqref{eq:Hdq_rot}] contains only single-qubit terms, its digital implementation requires only single-qubit quantum logic gates.  Figure~\ref{fig2} shows the circuit diagram for a six-qubit realization, where the Hamiltonian evolution is transpiled into the native IBMQ gate set. The qubits in this diagram are labeled as $q_i$ with $i \in [0,\,5]$, and their corresponding physical qubit indices on the selected backend are also indicated. In this circuit, the native gates acting on the $i$th qubit include the $z$-rotation $R_z(\theta) = e^{-i\frac{\theta}{2}\sigma_{iz}}$ with rotation angle $\theta$, and the $\mathrm{SX}$ gate, $\sqrt{X} = e^{-i\frac{\pi}{4}\sigma_{ix}}$. On IBMQ hardware, $R_z$ rotations are implemented virtually through frame updates and therefore require no physical gate execution. Consequently, the only physical operations applied to each qubit are the two $\sqrt{X}$ gates shown in the circuit.

\begin{figure}[htb]
    \includegraphics[width=\columnwidth, clip]{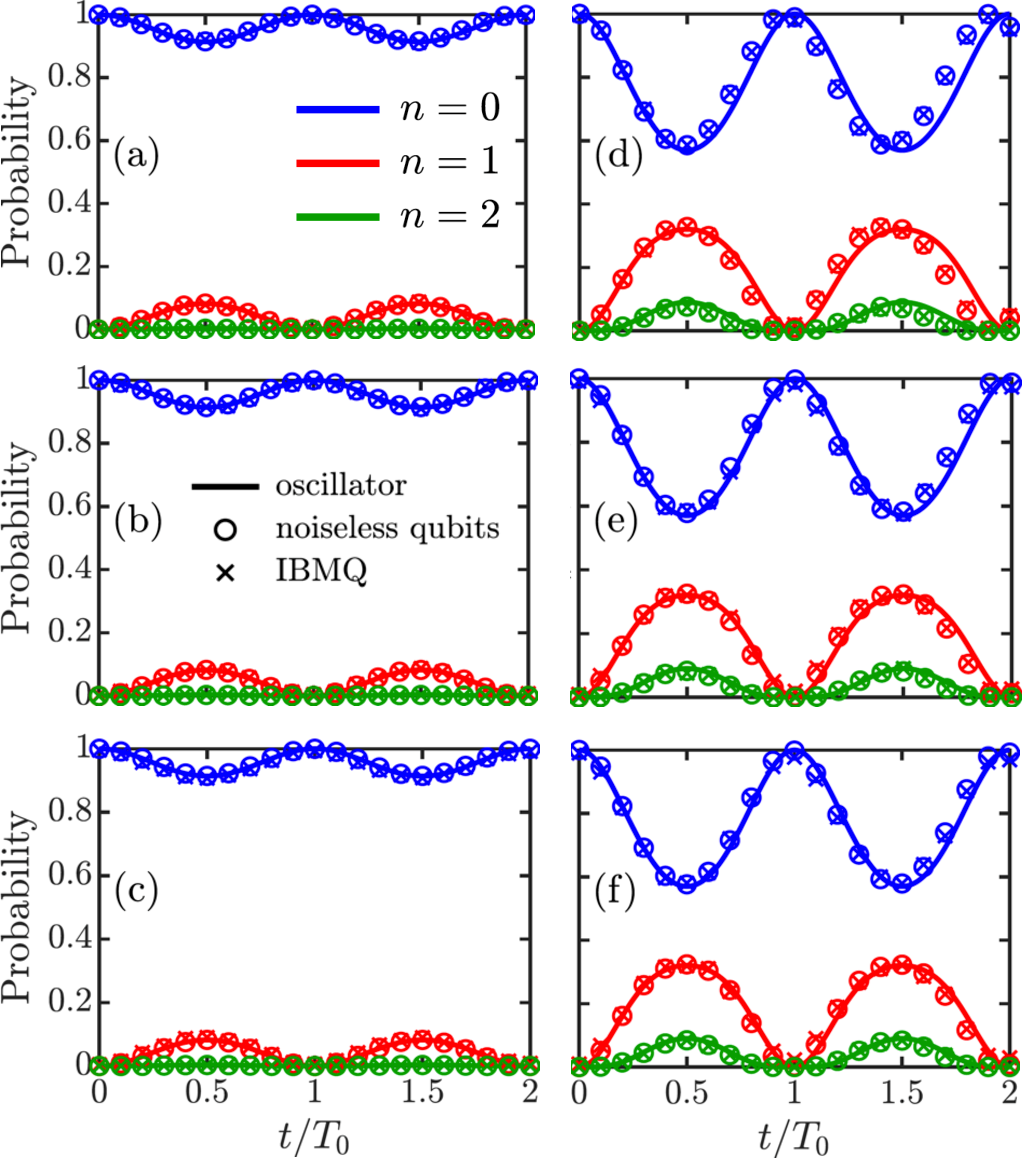}
    \caption{Probabilities of the ground state (blue, top), first excited state (red, middle), and second excited state (green, bottom) of the driven harmonic oscillator vs $t/T_{0}$. Solid lines: analytical results of the harmonic oscillator $P_{\mathrm{b}}^{(n)}$; circles: simulated results of the noiseless qubit ensemble $P_{\mathrm{c}}^{(n)}$; and crosses: results of noisy qubits from IBMQ $P_{\mathrm{q}}^{(n)}$. (a-c) $F = 0.3$ with $N = 3,\,5,\,7$ qubits; (d-f) $F = 0.75$ with $N = 6,\,11,\,14$ qubits, respectively. The detuning is $\Delta = 1$, and $T_{0} = \pi/\Omega$ is the oscillation period.}
    \label{fig3}
\end{figure}
In the simulations, we set the detuning $\Delta = 1$ and choose a driving amplitude $F < \Delta$ so that the mean excitation number remains below 1. Figure~\ref{fig3} shows the time evolution of the probabilities for the ground state $\ket{D_0^N}$ (corresponding to the oscillator ground state $\ket{0}$), the first excited state $\ket{D_1^N}$ ($\ket{1}$), and the second excited state $\ket{D_2^N}$ ($\ket{2}$). For each time $t$, the evolution is repeated 4000 times, and the probabilities are obtained by averaging the measurement outcomes. The differences between the probabilities for the harmonic oscillator and the noiseless qubit ensemble arise from the finite-$N$ approximation in the HP mapping of Eq.~\eqref{eq:HPT2}. Accordingly, these discrepancies are more pronounced for smaller ensemble sizes and gradually diminish as $N$ increases. As shown in Figs.~\ref{fig3}(c) and~\ref{fig3}(f), the differences become negligible when either the driving strength is small (e.g., $F = 0.3$) or the ensemble size is large (e.g., $N=14$). Meanwhile, the discrepancies between the results obtained from noiseless qubit ensemble and those from noisy quantum hardware originate from imperfections of the real device, and accumulate as the system size $N$ increases. Since this simulation involves only single-qubit operations with small gate errors, the dominant hardware-induced error arises from qubit readout inaccuracy. On IBMQ, the typical error rate of the $\sqrt{X}$ gates is below $8\times 10^{-4}$, whereas the readout error rate is approximately $4.5\times 10^{-3}$.  The device calibration data for the qubits used in these simulations are given in Table~\ref{tab1} of Appendix~\ref{sec:appa}.

\begin{figure}[htb]
    \includegraphics[width=\columnwidth, clip]{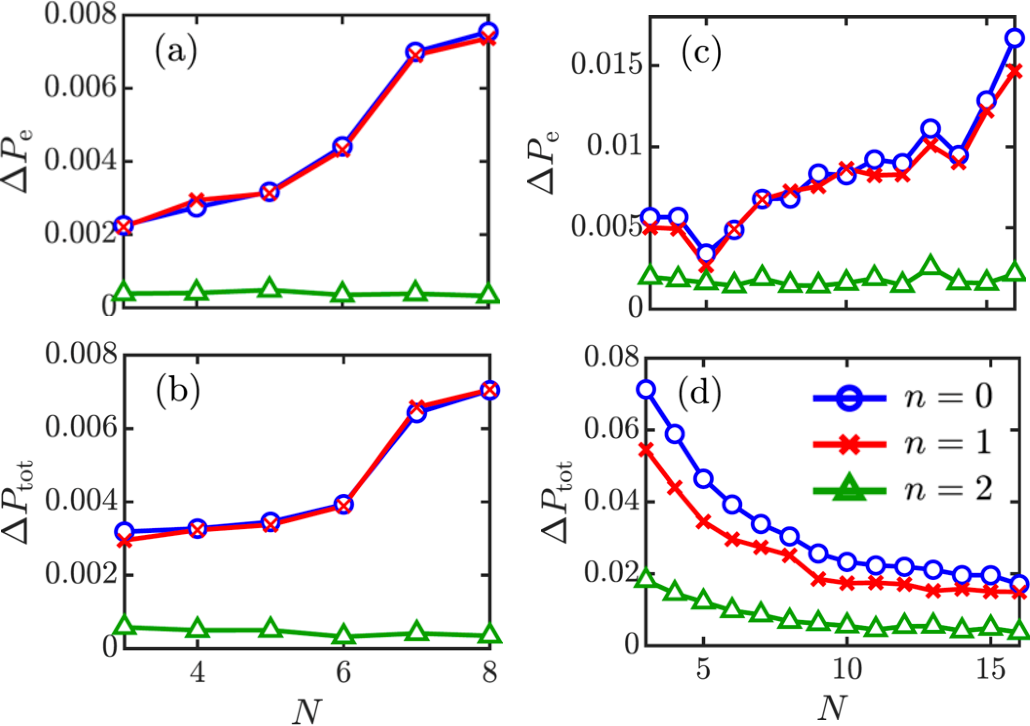}
    \caption{(a, c) Probability difference $\Delta P_{\mathrm{e}}^{(n)}$ vs qubit number $N$ for $F = 0.3$ and $F = 0.75$, respectively.  (b, d) Probability difference $\Delta P_{\mathrm{tot}}^{(n)}$ vs $N$ for $F = 0.3$ and $F = 0.75$, respectively.  Blue circles, for the ground state; red crosses, for the first excited state; and green triangles, for the second excited state. The lines are drawn to guide the eye, and $\Delta = 1$.}
    \label{fig4}
\end{figure}
To quantify the interplay between algorithmic errors arising from the HP mapping and hardware-induced errors in noisy devices, we first define the averaged probability difference that characterizes the effect of hardware-induced errors:
\begin{equation}
    \Delta P_{\mathrm{e}}^{(n)} 
    = \frac{1}{M}\sum_{j=1}^{M} 
      \left| P_{\mathrm{c}}^{(n)}(t_j) - P_{\mathrm{q}}^{(n)}(t_j) \right|,
    \label{eq:Pe}
\end{equation}
where $P_{\mathrm{c}}^{(n)}$ denotes the probability of the state $\ket{D_n^N}$ given in Eq.~\eqref{eq:PupN} for noiseless qubits, $P_{\mathrm{q}}^{(n)}$ is the corresponding probability obtained from noisy quantum hardware, $t_j$ is the $j$th time point, and $M$ is the total number of sampled time points during the evolution. For comparison, we define the average probability difference between the exact results for the bosonic mode $P_{\mathrm{b}}^{(n)}$ given in Eq.~\eqref{eq:Pn} and results obtained from the HP transformation using noiseless qubits:
\begin{equation}
    \Delta P_{\mathrm{a}}^{(n)} 
    = \frac{1}{M}\sum_{j=1}^{M} 
      \left| P_{\mathrm{b}}^{(n)}(t_j) - P_{\mathrm{c}}^{(n)}(t_j) \right|,
    \label{eq:Pa}
\end{equation}
which characterizes the algorithmic error associated with the HP mapping. The averaged probability difference 
\begin{equation}
    \Delta P_{\mathrm{tot}}^{(n)} 
    = \frac{1}{M}\sum_{j=1}^{M} 
      \left| P_{\mathrm{b}}^{(n)}(t_j) - P_{\mathrm{q}}^{(n)}(t_j) \right|,
    \label{eq:Ptot}
\end{equation}
captures the combined effect of the algorithmic and hardware-induced errors. 

Figure~\ref{fig4} shows the hardware-induced error $\Delta P_{\mathrm{e}}^{(n)}$ and the combined error $\Delta P_{\mathrm{tot}}^{(n)}$ as functions of the number of qubits $N$ for $F = 0.3$ and $F = 0.75$. As $N$ increases, $\Delta P_{\mathrm{e}}^{(n)}$ grows monotonically, reflecting the accumulation of gate and measurement errors with increasing system size.  In contrast, the algorithmic error decreases with increasing $N$ and vanishes in the limit $N \gg n$. Consequently, the combined deviation $\Delta P_{\mathrm{tot}}^{(n)}$ can either increase [Fig.~\ref{fig4}(b)] or decrease [Fig.~\ref{fig4}(d)] with $N$, depending on the competition between algorithmic and hardware-induced errors. For the weaker driving strength $F = 0.3$, the algorithmic error remains negligible due to the small excitation number, and $\Delta P_{\mathrm{tot}}^{(n)}$ is dominated entirely by hardware-induced errors. For $F = 0.75$, however, the algorithmic error becomes significant.  These results demonstrate that increasing the system size does not necessarily improve simulation accuracy, highlighting the trade-off between algorithmic approximations and hardware noise in near-term quantum simulations. We note that the circuit depth, and hence the hardware execution time, remains fixed for all evolution times $t$ and the accumulated hardware error does not systematically increase with $t$. Therefore, it is appropriate to quantify the simulation using the averaged probability differences, which provide a reliable measure by incorporating information from multiple time points in the evolution. 

\begin{figure}[t]
    \includegraphics[width=\columnwidth, clip]{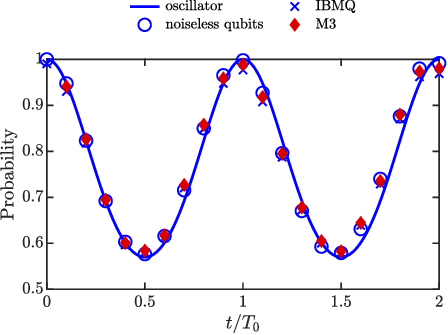}
    \caption{Probabilities of the ground state of the driven oscillator vs $t/T_{0}$. Solid lines, analytical results of the harmonic oscillator $P_{\mathrm{b}}^{(0)}$; circles, simulated results of the noiseless qubit ensemble $P_{\mathrm{c}}^{(0)}$; crosses, results of noisy qubits from IBMQ $P_{\mathrm{q}}^{(0)}$; and red diamonds, M3-mitigated results $P_{\mathrm{QEM}}^{(0)}$. Here $F = 0.75$ and $\Delta=1$ with $N = 14$ qubits.}
    \label{fig5}
\end{figure}
To reduce the effect of readout errors in the above simulation, we apply the matrix-free measurement mitigation (M3) technique described in Appendix~\ref{ssec:M3}. The ground-state probabilities for the simulated driven oscillator with $F=0.75$ and $N=14$ qubits, including noisy qubits and error-mitigated results, are presented in Fig.~\ref{fig5}. Compared with the unmitigated hardware results, the M3-corrected data exhibit improved agreement with the noiseless qubit results throughout the time evolution. The averaged difference between the noiseless qubit data and the noisy qubit data decreases from $9.5\times 10^{-3}$ to $6.3\times 10^{-3}$. This improvement further supports our analysis that readout errors are a dominant source of hardware error in the simulation of the driven harmonic oscillator, whose quantum circuit consists solely of single-qubit gates.

\subsection{Circuit Implementation of Squeezing \label{ssec:squeezing_implement}}
To simulate squeezing, we employ the first-order Trotter expansion (see Appendix~\ref{sec:TrotterMethod}) on the operator $S(r, \phi)$ given in Eq.~(\ref{eq:squeezing0}), with $S(r,\phi)\approx \left[ \prod_{i,j=1}^{N} S_{ij}(r_0,\phi) \right]^{K_T}\!\!$, where $K_T$ is the number of Trotter steps, $r_0=\frac{r}{2NK_T}$, and
\begin{equation}
    S_{ij}(r_0,\phi)=
    \exp\!\left[
    r_0
    \left(
    e^{-2i\phi}\sigma_{i-}\sigma_{j-}
    -
    e^{2i\phi}\sigma_{i+}\sigma_{j+}
    \right)
    \right].
    \label{eq:Sij}
\end{equation}
For the squeezing phase $\phi=0$, each two-qubit operator $S_{ij}$ can be expressed as an $R_{XX-YY}$ interaction with
\begin{equation}
R_{XX-YY}(\theta,\beta)=
\begin{pmatrix}
\cos\frac{\theta}{2} & 0 & 0 &
-i\sin\frac{\theta}{2}e^{-i\beta} \\
0 & 1 & 0 & 0 \\
0 & 0 & 1 & 0 \\
-i\sin\frac{\theta}{2}e^{i\beta} & 0 & 0 &
\cos\frac{\theta}{2}
\end{pmatrix},
\end{equation}
$\theta=2r_0$, and $\beta=\pi/2$. This interaction can be transpiled by Qiskit into the native gate set of IBM Kingston, consisting of CZ gates and single-qubit rotations.

\begin{figure}
    \centering
    \includegraphics[width=\columnwidth, clip]{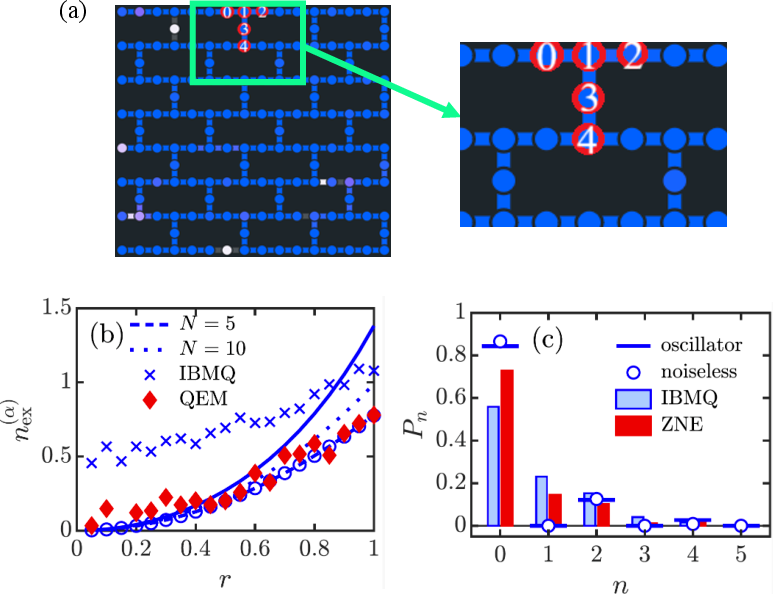}
    \caption{(a) Layout for the qubits $q_0$--$q_4$ in the digital simulation of squeezing on IBM Kingston. (b) Mean excitation number $n_{\rm ex}^{(\alpha)}$ vs the squeezing parameter $r$ and (c) excitation-number distribution $P_{\alpha}^{(n)}$ at $r = 0.6$, with $\alpha={\rm b}, {\rm c}, {\rm q}, {\rm QEM}, {\rm ZNE}$ representing the oscillator (solid lines), noiseless qubits [dashed (dotted) line for Eq.~\eqref{eq:squeezing0} with $N = 5$ ($N=10$), and circles for results with a single Trotter step $K_T=1$], IBMQ (crosses and light bars), and error-mitigated results [red diamonds for full QEM (M3+ZNE) and red bars for ZNE only], respectively.}
    \label{fig6}
\end{figure}
We employ a five-qubit ensemble to demonstrate the digital simulation of the squeezing operator. Since implementing each $S_{ij}$ requires entangling gates and the squeezing operation involves all-to-all qubit-qubit interactions, we select physical qubits that are as closely connected as possible within the connectivity of IBM Kingston to minimize the circuit depth and accumulated gate errors. The corresponding qubit layout is shown in Fig.~\ref{fig6}(a) with the qubit data given in Tables~\ref{tab2} and \ref{tab3} of Appendix~\ref{sec:appa}. With this layout, each Trotter step includes 35 CZ gates after transpilation. To characterize the generated squeezed state, we measure the average number of excitations in the qubit ensemble and compare it with the average boson number $n_{\rm ex}^{({\rm b})}$. Figure~\ref{fig6}(b) shows $n_{\rm ex}^{(\alpha)}$ as a function of the squeezing parameter $r$, where $\alpha = {\rm b, c, q, QEM}$ denotes the ideal bosonic mode, noiseless qubit ensemble, noisy qubit ensemble, and full error-mitigated (M3+ZNE) results, respectively. The result for noiseless qubits simulated using Qiskit Aer with a single Trotter step ($K_T=1$, circles) agrees with the exact evolution under Eq.~(12) for $N=5$ (dashed curve) to within $0.005$ over the entire range of $r$, indicating that the Trotter error is negligible. The noiseless results are close to the oscillator results (solid curve) for small and moderate squeezing strengths, but show a large discrepancy at large $r$. This discrepancy arises from the finite size of the qubit ensemble, i.e., the large-$N$ approximation in Eq.~(3) and the truncation of the Hilbert space at $n=N$, both of which become more pronounced as $\langle n\rangle$ increases. The deviation decreases with $N$, as illustrated by the result for $N=10$ (dotted curve). The hardware results obtained on IBM Kingston follow the same overall trend but exhibit a noticeable baseline deviation from the noiseless qubit results across the entire range of $r$. We employ M3 to reduce readout errors and ZNE to mitigate errors associated with the two-qubit gates (see Appendix~\ref{ssec:ZNE}). The error-mitigated results exhibit significant improvement and show much better agreement with the noiseless qubit results. To further characterize the generated states, we show in Fig.~\ref{fig6}(c) the distribution of the excitation number $P_{\alpha}^{(n)}$ at $r=0.6$, where $\alpha={\rm b, c, q, ZNE}$, and ${\rm ZNE}$ denotes the hardware results mitigated with ZNE. For an ideal squeezed vacuum, only even excitation numbers are populated, with $P_{\rm b}^{(2n)} = \frac{(2n)!}{2^{2n}(n!)^2}\frac{\tanh^{2n}r}{\cosh r}$ and $P_{\rm b}^{(2n+1)}=0$; the noiseless qubit results $P_{\rm c}^{(n)}$ reproduce this parity structure. In the IBMQ simulation, hardware errors populate the odd excitation numbers, mainly $n=1$. Nevertheless, the parity of the error-mitigated distribution, $\sum_n(-1)^nP_{\rm ZNE}^{(n)} = 0.69$, exceeds that of a coherent state ($0.42$) or a thermal state ($0.53$) with the same mean excitation number ($0.44$), reflecting the pairwise creation of excitations by the squeezing operator. Note that because only ZNE is applied here, the mean excitation number of the mitigated distribution ($0.44$) is slightly larger than $n_{\rm ex}^{\rm (QEM)}=0.39$ in Fig.~\ref{fig6}(b), due to uncorrected readout errors. We also note that measuring the excitation number does not distinguish symmetric Dicke states from other states with the same number of excitations. Hardware errors can induce leakage out of the symmetric sector, providing an additional source of deviation from the ideal bosonic dynamics. Meanwhile, error mitigation improves the measured distribution but does not restore the permutation symmetry.

\section{Digital Simulation of the JC Model \label{sec:JCM}}
\subsection{JC Model and HP Transformation \label{ssec:JCMModel}}
The JC model describes the interaction between a single bosonic mode (cavity) and a quantum two-level system (qubit), with the Hamiltonian
\begin{equation}
    H = \omega_0 a^{\dagger}a 
      + \frac{\omega_z}{2}\tau_z
      + g\left(a\,\tau_{+} + a^{\dagger}\tau_{-}\right),
    \label{eq:HJC}
\end{equation}
where $\tau_z$ and $\tau_{\pm}$ are the Pauli operators of the qubit,  $\omega_z$ is the qubit energy splitting, and $g$ is the coupling strength between the qubit and the cavity mode.  The JC model plays a central role in quantum optics and quantum information, exhibiting rich coherent dynamics that have been implemented in both atomic and solid-state quantum systems~\cite{BlaisPRA2004}. 

For the initial state $\ket{0,\uparrow_{\tau}}$, where the qubit is in the spin-up state and the cavity is in its vacuum state, the system undergoes vacuum Rabi oscillations between the states $\ket{0,\uparrow_{\tau}}$ and $\ket{1,\downarrow_{\tau}}$.  This dynamics corresponds to the coherent exchange of a single excitation between the qubit and the cavity. The probability that the system remains in the state $\ket{0,\uparrow_{\tau}}$ at time $t$ is
\begin{equation}
    P_{\mathrm{b}}^{(0, \uparrow)}(t)
    = 1 -  \frac{g^2}{\Omega_R^2} \sin^2(\Omega_R t) 
    \label{eq:PupJC}
\end{equation}
with the Rabi frequency $\Omega_R = \sqrt{g^2 + \Delta^2/4}$ and the qubit-cavity detuning $\Delta = \omega_z - \omega_0$. For the initial state $\ket{2,\downarrow_\tau}$, the dynamics are confined to the two-dimensional subspace $\{\ket{2,\downarrow_\tau},\ket{1,\uparrow_\tau}\}$. The probability of finding the system in the state $\ket{2,\downarrow}$ at time $t$ can be derived similarly as  
\begin{equation}
P_{\rm b}^{(2,\downarrow)}(t) =
1- \frac{g_2^2}{\Omega_{R2}^2} \sin^2\!\left(
\Omega_{R2} t
\right), 
\label{eq:PdownJC2}
\end{equation}
with the corresponding Rabi frequency $\Omega_{R2} = \sqrt{g_2^2+\Delta^2/4}$ and $g_2 = \sqrt{2} g$.

Using the HP transformation in Eq.~\eqref{eq:HPT2}, the cavity mode can be mapped onto a qubit ensemble. The JC Hamiltonian in Eq.~\eqref{eq:HJC} is mapped to
\begin{equation}
    H = \frac{\omega_0}{2}\sum_{i=1}^{N} \sigma_{iz} 
      + \frac{\omega_z}{2}\tau_{z}
      + \frac{g}{2\sqrt{N}}\sum_{i=1}^{N}
        \left(\sigma_{ix}\tau_{x} + \sigma_{iy}\tau_{y}\right),
    \label{eq:HJCq}
\end{equation}
where the constant term $\tfrac{N}{2}\omega_0$ has been omitted. In contrast to the driven harmonic oscillator case, the JC Hamiltonian contains interactions between the cavity qubits (described by the $\sigma_i$ operators) and the qubit coupled to the cavity mode (described by the $\tau$ operators). These interactions can be expressed in the form of $XX$ and $YY$ couplings. Implementing this Hamiltonian therefore requires two-qubit quantum logic gates. Moreover, the interaction terms do not commute. For example, the interaction between cavity qubit $\sigma_1$ and $\tau$ does not commute with the interaction between cavity qubit $\sigma_2$ and $\tau$. To realize the dynamics of this Hamiltonian on IBMQ hardware, we employ two approaches: the Suzuki-Trotter decomposition and a synthesized unitary approach, as discussed below. 

For a cavity represented by an ensemble of $N$ qubits, the vacuum state $|0\rangle$ is mapped to the Dicke state $|D_0^N\rangle = |\downarrow\downarrow\cdots\downarrow\rangle$, in which all qubits are in the spin-down state, while the one-photon Fock state $|1\rangle$ is mapped to the symmetric Dicke state $|D_1^N\rangle = \frac{1}{\sqrt{N}} \sum_{i=1}^{N}
|\downarrow\cdots\uparrow_i\cdots\downarrow\rangle$. Starting from the state $\ket{D_0^N,\uparrow_{\tau}}$, the dynamics are restricted to the single-excitation subspace spanned by $\{\ket{D_0^N,\uparrow_{\tau}},\ket{D_1^N,\downarrow_{\tau}}\}$, with $J_+|D_0^N\rangle=\sqrt{N}\,|D_1^N\rangle$. We find that the probability of the state $\ket{D_0^N,\uparrow_{\tau}}$ is identical to the exact JC result, $P_{\mathrm{c}}^{(0, \uparrow)}(t) \equiv P_{\mathrm{b}}^{(0, \uparrow)}(t)$, given by Eq.~(\ref{eq:PupJC}). Hence, the HP mapping exactly reproduces the vacuum Rabi oscillation of the JC model in this case, independent of the number of qubits $N$. For the initial state $\ket{D_2^N,\downarrow_\tau}$, the dynamics are confined to the two-dimensional subspace $\{\ket{D_2^N,\downarrow_\tau}, \ket{D_1^N,\uparrow_\tau}\}$. It can be shown that $J_{-} \ket{D_2^N} = \sqrt{2(N-1)}\,\ket{D_1^N}$, which leads to the effective coupling between these two states $g_{N,2} = g\sqrt{\frac{2(N-1)}{N}}$. The probability of finding the system in the state $\ket{D_2^N,\downarrow}$ at time $t$ is therefore
\begin{equation}
P_{\rm c}^{(2,\downarrow)}(t) = 1- 
\frac{g_{N,2}^2}{\Omega_{R2,c}^2} \sin^2\left(
\Omega_{R2,c} t \right),
\label{eq:PdownJC2N}
\end{equation}
with the Rabi frequency $\Omega_{R2,c} = \sqrt{g_{N,2}^2+\Delta^2/4}$.
In the limit $N\rightarrow\infty$, the effective coupling $g_{N,2}\rightarrow g_2$, recovering the exact JC result in Eq.~(\ref{eq:PdownJC2}).

\subsection{Circuit Implementation of Single-Excitation Dynamics \label{ssec:1excitation}}
We first study the digital simulation of the single-excitation dynamics of the JC model, where the HP transformation exactly reproduces the vacuum Rabi oscillation, as discussed in Sec.~\ref{ssec:JCMModel}. We implement the dynamics using both the Suzuki-Trotter decomposition and the synthesized unitary approach, and conduct a diagnostic analysis of the effects of hardware noise on the simulation.

\begin{figure}[t]
    \centering
    \includegraphics[width=\columnwidth, clip]{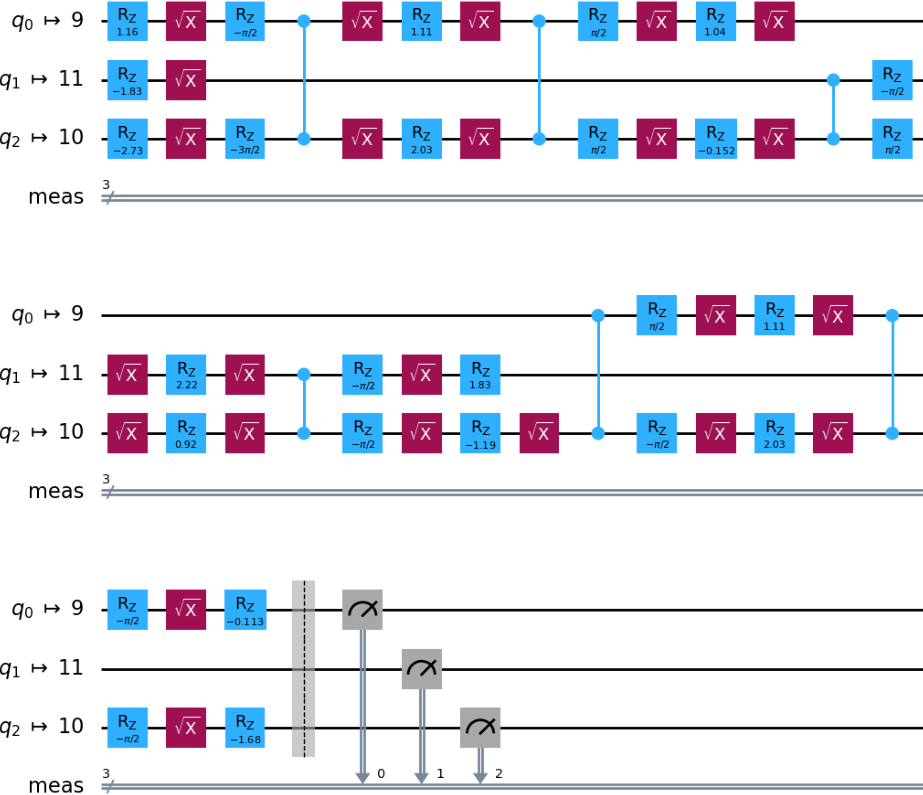}
    \caption{Circuit diagram of a single Suzuki-Trotter step (up to a global phase of $\pi$) for the digital simulation of the JC model using $N=2$ qubits to represent the cavity mode ($q_0=\sigma_1$, $q_1=\sigma_2$) and one qubit to couple with the cavity mode ($q_2=\tau$). The circuit is transpiled into native IBMQ gates followed by qubit measurements. The solid bars connecting pairs of qubits denote the CZ gates.  The phases of the $R_z$ gates correspond to the parameters $\omega_0 = \omega_z = 1.0$, $g = 0.1$, and the evolution time $t = \pi$.}   
    \label{fig7}
\end{figure}
\subsubsection{Trotterization and Implementation \label{sssec:4B1}}
We adopt the Suzuki-Trotter decomposition (see Appendix~\ref{sec:TrotterMethod}) to implement the JC Hamiltonian on IBMQ hardware. For $\omega_0 = \omega_z = 1$ and $H_0 = \frac{\omega_0}{2}(\sum_{i=1}^{N} \sigma_{iz} + \tau_{z})$, the JC Hamiltonian in the rotating frame defined by $H_0$ reduces to the interaction term given in Eq.~\eqref{eq:HJCq}. We first use $N = 2$ qubits to represent the cavity mode. The corresponding Suzuki-Trotter decomposition of the rotating-frame Hamiltonian $H_{\mathrm{rot}}$ takes the form:
\begin{equation}
    e^{-i t H_{\mathrm{rot}}} 
    \approx 
    \left(
        e^{-i H_1 \frac{t}{2K_T}} 
        e^{-i H_2 \frac{t}{K_T}} 
        e^{-i H_1 \frac{t}{2K_T}}
    \right)^{K_T},
    \label{eq:JC_Trotter}
\end{equation}
where $H_1$ and $H_2$ denote the noncommuting components of $H_{\mathrm{rot}}$: $H_1$ describes the coupling between the cavity qubit $\sigma_1$ and the qubit $\tau$, while $H_2$ accounts for the interaction between the cavity qubit $\sigma_2$ and $\tau$. Each Suzuki-Trotter step can be implemented using the following circuit:
\\[8pt]
{\begin{adjustbox}{width=0.48\textwidth}
\begin{quantikz}
\lstick[2]{{\large Cavity} \\ {\large in JC}} 
& \gate{R_Z} & \qw & \qw & \gate[3]{R_{XX}} & \gate[3]{R_{YY}} & \qw & \qw & \gate{R_Z} & \qw \\ 
& \gate{R_Z} & \gate[2]{R_{XX}} & \gate[2]{R_{YY}} & \qw & \qw & \gate[2]{R_{XX}} & \gate[2]{R_{YY}} & \gate{R_Z} & \qw \\
\lstick{{\large Qubit} \\ {\large in JC}} 
& \gate{R_Z} & \qw & \qw & \qw & \qw & \qw & \qw & \gate{R_Z} & \qw
\end{quantikz}
\end{adjustbox}}
\\[8pt]
which includes single-qubit $R_Z$ gates and the two-qubit gates $R_{XX}(\theta) = e^{-i \frac{\theta}{2} X_1 \otimes X_2}$ and $R_{YY}(\theta) = e^{-i \frac{\theta}{2} Y_1 \otimes Y_2}$. When transpiled into the native IBMQ basis set, the combination of one $R_{XX}$ and one $R_{YY}$ operation can be implemented using two Controlled-Z (CZ) gates, together with several single-qubit rotations denoted by $U$:
\\[8pt]
{\begin{adjustbox}{width=0.48\textwidth}
\begin{quantikz}
    & \gate[2]{R_{XX}} & \gate[2]{R_{YY}} & \qw \\
    & & \qw & \qw
\end{quantikz}
\;\; = \;\;
\begin{quantikz}
    & \gate{U} & \gate[2]{R_{\mathrm{CZ}}} & \gate{U} & \gate[2]{R_{\mathrm{CZ}}} & \gate{U} & \qw \\
    & \gate{U} & & \gate{U} & & \gate{U} & \qw
\end{quantikz}
\end{adjustbox}}
\\[8pt]
The complete circuit diagram of a single Suzuki-Trotter step includes six CZ gates and is shown in Fig.~\ref{fig7}, where all single-qubit operations are decomposed into $R_Z$ and $\sqrt{X}$ gates. For $K_T$ Trotter steps, circuit transpilation reduces the total number of CZ gates to $4K_T+2$, which increases linearly with $K_T$.

\begin{figure}[t]
    \includegraphics[width=\columnwidth, clip]{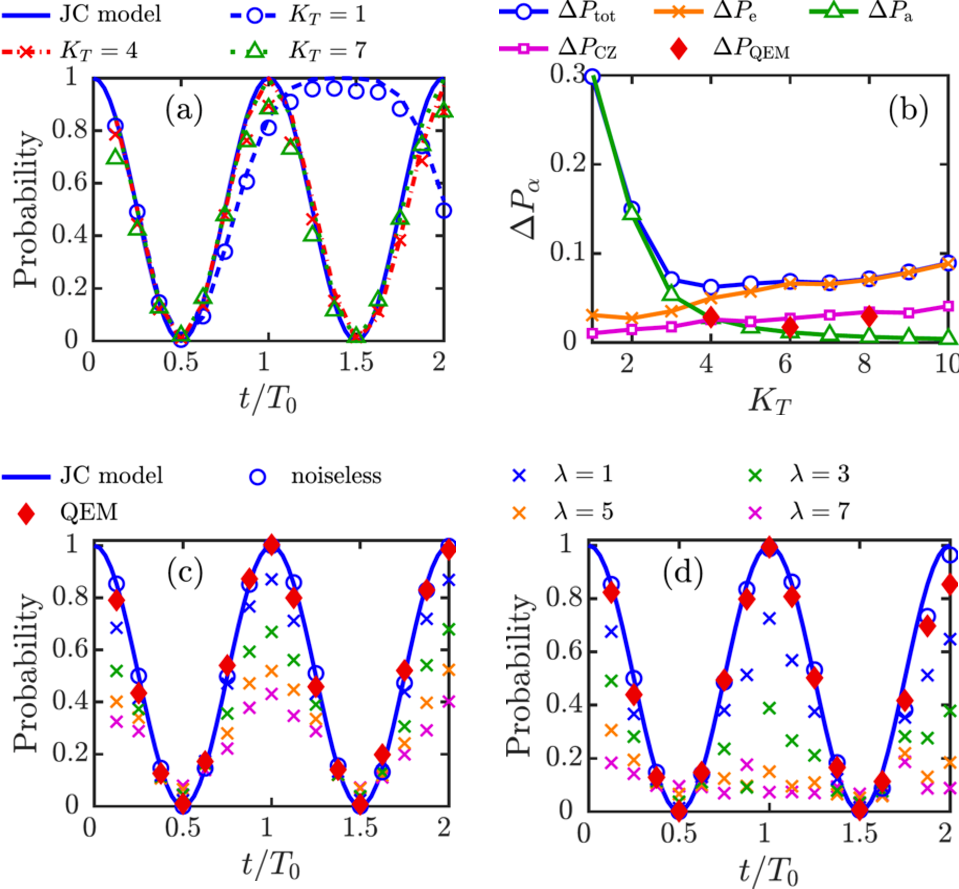} 
    \caption{(a) Probability of $\ket{0,\uparrow_{\tau}}$ vs $t/T_0$ from Trotterization. Blue dashed line (circles): $K_T = 1$ for noiseless (noisy) qubits; red dash-dotted line (crosses): $K_T = 4$ for noiseless (noisy) qubits; and green dotted line (triangles): $K_T = 7$ for noiseless (noisy) qubits. (b) Probability difference $\Delta P_{\alpha}$ vs $K_T$ with $\alpha = \mathrm{tot, e, a, QEM}$. Blue circles, $\Delta P_{\mathrm{tot}}$; orange crosses, $\Delta P_{\mathrm{e}}$; green triangles, $\Delta P_{\mathrm{a}}$; red diamonds, $\Delta P_{\mathrm{QEM}}$; and magenta squares, $\Delta P_{\mathrm{CZ}}$ for the CZ-gate-only circuit in Sec.~\ref{sssec:4B3}. Solid lines are drawn to guide the eye. (c, d) Probability of $\ket{0,\uparrow_{\tau}}$ with QEM vs $t/T_0$ for $N=2$ qubits with $K_T=8$ and $N=3$ qubits with $K_T=4$, respectively. Crosses: results for noisy qubits with $\lambda=1, 3, 5, 7$ (blue, green, orange, magenta) for ZNE; and red diamonds: QEM results. Here $\omega_0=\omega_z=1$, $g=0.1$, and $T_0 = \pi/\Omega_R$. In (a), (c), and (d), the blue solid lines are analytical results for the JC model.}
    \label{fig8}
\end{figure}
To simulate the vacuum Rabi oscillation between the states $\ket{0,\uparrow_{\tau}}$ and $\ket{1,\downarrow_{\tau}}$, the qubit in the JC model is initialized in the spin-up state, and the $N$ qubits representing the cavity mode are initialized in the spin-down state.  We then apply the Trotterized circuit for a given evolution time $t$, spanning up to two full oscillation periods, followed by qubit measurements to obtain the probability of the state $\ket{0,\uparrow_{\tau}}$.  For each value of $t$, the circuit is executed 4{,}000 times to estimate the probability. We measure this probability for $K_T = 1$--$7$ Trotter steps to investigate the interplay between algorithmic and hardware-induced errors, analogous to the analysis performed in Sec.~\ref{sec:dho}. For this circuit, a major source of hardware error is the CZ gates with error rates on the order of $3.3\times 10^{-2}$ (see Table~\ref{tab4} of Appendix~\ref{sec:appa}).  

Figure~\ref{fig8}(a) shows the simulated probability of the state $\ket{0, \uparrow_{\tau}}$ as a function of $t/T_0$, where $T_0$ denotes the period of the Rabi oscillation. In Fig.~\ref{fig8}(b), we plot the probability differences $\Delta P_{\mathrm{\alpha}}$ ( $\alpha = {\rm tot, e, a}$) defined in Sec.~\ref{ssec:dho_implement} as functions of the number of Trotter steps $K_T$. In this case, the probability difference due to algorithmic error $\Delta P_{\mathrm{a}}$ arises solely from the finite number of Trotter steps. Our results show that $\Delta P_{\mathrm{e}}$ generally increases with $K_T$ as the circuit depth (and thus accumulated gate error) grows, whereas $\Delta P_{\mathrm{a}}$ decreases with $K_T$ as the Trotter approximation improves. The total probability difference $\Delta P_{\mathrm{tot}}$ exhibits an optimal minimum around $K_T \approx 4$, reflecting the interplay between algorithmic and hardware-induced errors.

We further perform the simulation for $N=3$ qubits to examine the effect of increasing the size of the qubit ensemble. For $N=3$, a single Trotter step requires ten CZ gates; for $K_T$ Trotter steps, circuit transpilation reduces the total number of CZ gates to $8K_T+2$. Hence, for $K_T=4$, the $N=3$ circuit contains 34 CZ gates in total, the same as the $N=2$ circuit with $K_T=8$. Nevertheless, the $N=3$ simulation exhibits a somewhat larger hardware error than the $N=2$ simulation. As $K_T$ increases, the larger number of entangling gates and greater circuit depth are expected to further increase the accumulated hardware errors, consistent with the behavior observed for $N=2$. This behavior illustrates a practical limitation of extending the HP-based simulation to larger qubit ensembles on current quantum hardware. The qubit data in the simulation can be found in Table~\ref{tab6} of Appendix~\ref{sec:appa}.

To improve the results from the quantum hardware, we apply QEM to the above simulations. Figures~\ref{fig8}(c) and (d) show the vacuum Rabi oscillations after applying M3 measurement mitigation and ZNE for $N=2$ with $K_T=8$ Trotter steps and $N=3$ with $K_T=4$ Trotter steps, respectively. For ZNE, data are collected for $0$--$3$ folds of the unitary evolution (i.e., $\lambda=1, 3, 5, 7$), corresponding to progressively increased noise levels, and are extrapolated to the zero-noise limit as described in Appendix~\ref{ssec:ZNE}. Compared with the unmitigated results, the error-mitigated data show substantially improved agreement with the noiseless qubit simulations. To quantify the impact of QEM, we define the averaged probability difference $\Delta P_{\rm QEM}$ by replacing the raw hardware results in Eq.~(\ref{eq:Ptot}) with the error-mitigated results. The averaged probability difference is reduced from the unmitigated $\Delta P_{\rm tot}=0.071$ to $\Delta P_{\rm QEM}=0.029$ for $N=2$ [see Fig.~\ref{fig8}(b)], and from $\Delta P_{\rm tot}=0.131$ to $\Delta P_{\rm QEM}=0.042$ for $N=3$. As shown by the red diamonds in Fig.~\ref{fig8}(b), the magnitudes of $\Delta P_{\rm QEM}$ are close to those of $\Delta P_{\rm a}$, particularly at $K_T = 4$, demonstrating that the combined use of M3 and ZNE effectively reduces readout and entangling-gate errors in the hardware implementation of the JC dynamics. In general, our results indicate that hardware-induced noise is a major limitation for extending the HP-based simulations to more complex scenarios on current quantum hardware, while QEM can partially mitigate this limitation.

\subsubsection{Synthesized Unitary Approach and Implementation \label{sssec:4B2}}
In this section, we employ the synthesized unitary approach with a fixed quantum circuit layout, where a constant number of two-qubit gates are interleaved with single-qubit rotations containing tunable parameters to simulate the unitary evolution of the JC model~\cite{Goubault2019synthesizing, Ashhab2022numerical, Ashhab2024quantum, BablerQuantum2023}. The parameters of the single-qubit gates are optimized using classical simulations to best reproduce the desired target unitary evolution. Each single-qubit gate is represented by a general single-qubit unitary operator:
\begin{equation}
    U(\alpha, \beta, \gamma) =
    \begin{pmatrix}
        \cos\frac{\beta}{2} & -e^{i\gamma}\sin\frac{\beta}{2} \\ 
        e^{i\alpha}\sin\frac{\beta}{2} & e^{i(\alpha+\gamma)}\cos\frac{\beta}{2}
    \end{pmatrix},
\end{equation}
where the parameters $\alpha$, $\beta$, and $\gamma$ are numerically optimized following the standard $U3$-gate convention used in Qiskit. The cost function for the optimization is defined as $C = 1 - F$, where
\begin{equation}
    F = \left|\frac{1}{2^{N+1}} \mathrm{Tr}\!\left(U_f^{\dagger} U_T\right)\right|^2
    \label{fidelity}
\end{equation}
is the fidelity between the synthesized unitary transformation $U_f$ and the target unitary transformation $U_T$ at a given time $t$ with $N + 1$ being the total number of qubits in the simulated JC model. Using the limited-memory Broyden-Fletcher-Goldfarb-Shanno (L-BFGS) algorithm for optimization, we determine the optimal set of single-qubit rotation parameters that maximize $F$. In the optimization, the entangling gates are chosen to be the CZ gate to ensure compatibility with IBM Kingston. Compared with the Suzuki-Trotter expansion described in Sec.~\ref{sssec:4B1}, the synthesized unitary approach requires only a fixed, small number of entangling gates and can be implemented with a circuit depth comparable to that of a single Trotter step. In this way, errors arising from two-qubit operations are reduced, allowing significantly higher fidelity to be achieved. We note that this method is limited to systems with a small number of qubits, since the classical computational cost of optimizing the circuit parameters increases exponentially with system size and the optimization needs to be conducted for each given evolution time $t$.

To simulate the JC Hamiltonian for $N = 2$, we test circuit layouts containing either four or six CZ gates. The four-CZ-gate circuit employs the same number of two-qubit operations as a standard first-order (non-symmetrized) Trotter decomposition. We examine three distinct four-CZ-gate configurations:
\\[8pt]
{\begin{adjustbox}{width=0.48\textwidth}
\begin{quantikz}
\lstick[2]{{\large Cavity} \\ {\large in JC}}  
& \gate{U} & \qw & \qw & \qw & \qw 
    & \gate[3]{R_{\mathrm{CZ}}} & \gate{U} & \gate[3]{R_{\mathrm{CZ}}} & \gate{U} & \qw \\
& \gate{U} & \gate[2]{R_{\mathrm{CZ}}} & \gate{U} & \gate[2]{R_{\mathrm{CZ}}} & \gate{U} 
    & & \qw & & \qw & \qw \\
\lstick{{\large Qubit} \\ {\large in JC}} 
& \gate{U} & & \gate{U} & & \gate{U} 
    & & \gate{U} & \qw & \gate{U} & \qw
\end{quantikz}
\end{adjustbox}}
\\[8pt]
{\begin{adjustbox}{width=0.48\textwidth}
\begin{quantikz}
\lstick[2]{{\large Cavity} \\ {\large in JC}}  
& \gate{U} & \qw & \qw & \gate[3]{R_{\mathrm{CZ}}} & \gate{U} 
    & \qw & \qw & \gate[3]{R_{\mathrm{CZ}}} & \gate{U} & \qw \\
& \gate{U} & \gate[2]{R_{\mathrm{CZ}}} & \gate{U} & \qw & \qw 
    & \gate[2]{R_{\mathrm{CZ}}} & \gate{U} & \qw & \qw & \qw \\
\lstick{{\large Qubit} \\ {\large in JC}} 
& \gate{U} & & \gate{U} & & \gate{U} & & \gate{U} & \qw & \gate{U} & \qw
\end{quantikz}
\end{adjustbox}}
\\[8pt]
{\begin{adjustbox}{width=0.48\textwidth}
\begin{quantikz}
\lstick[2]{{\large Cavity} \\ {\large in JC}}  
& \gate{U} & \qw & \qw & \gate[3]{R_{\mathrm{CZ}}} & \gate{U} 
    & \gate[3]{R_{\mathrm{CZ}}} & \gate{U} & \qw & \qw & \qw \\
& \gate{U} & \gate[2]{R_{\mathrm{CZ}}} & \gate{U} & \qw & \qw 
    & \qw & \qw & \gate[2]{R_{\mathrm{CZ}}} & \gate{U} & \qw \\
\lstick{{\large Qubit} \\ {\large in JC}} 
& \gate{U} & & \gate{U} & & \gate{U} & & \gate{U} & & \gate{U} & \qw
\end{quantikz}
\end{adjustbox}}
\\[8pt]
These layouts are selected according to the following criteria: (1) each qubit representing the cavity mode participates in the same number of CZ gates; (2) every CZ gate couples one cavity-mode qubit with the qubit in the JC model; and (3) each layout is unique up to a reversal of the gate order or an exchange of the cavity-mode qubit indices, thereby eliminating redundant configurations. For each evolution time $t$, all three configurations are simulated, and the layout yielding the highest fidelity $F$ is selected. The six-CZ-gate layout employs the same number of two-qubit operations as the Suzuki-Trotter decomposition and takes the form:
\\[8pt]
{\begin{adjustbox}{width=0.48\textwidth}
\begin{quantikz}\label{3qbU}
\lstick[2]{{\large Cavity} \\ {\large in JC}}  
& \gate{U} & \qw & \qw & \qw & \qw & \gate[3]{R_{\mathrm{CZ}}} 
    & \gate{U} & \gate[3]{R_{\mathrm{CZ}}} & \gate{U} & \qw & \qw & \qw & \qw & \qw \\
& \gate{U} & \gate[2]{R_{\mathrm{CZ}}} & \gate{U} & \gate[2]{R_{\mathrm{CZ}}} & \gate{U} 
    & & \qw & & \qw & \gate[2]{R_{\mathrm{CZ}}} & \gate{U} & \gate[2]{R_{\mathrm{CZ}}} & \gate{U} & \qw \\
\lstick{{\large Qubit} \\ {\large in JC}}  
& \gate{U} & \qw & \gate{U} & \qw & \gate{U} & \qw 
    & \gate{U} & \qw & \gate{U} & \qw & \gate{U} & \qw & \gate{U} & \qw
\end{quantikz}
\end{adjustbox}}
\\[8pt]
This circuit layout is essentially identical to a single Suzuki-Trotter step for the JC model shown in Fig.~\ref{fig7} except for the rotation angles of the single-qubit gates. Using this layout, we achieve very high gate fidelities, with $1 - F < 10^{-5}$ for arbitrary evolution times~$t$.

\begin{figure}[hbt]
    \centering
    \includegraphics[width=\columnwidth, clip]{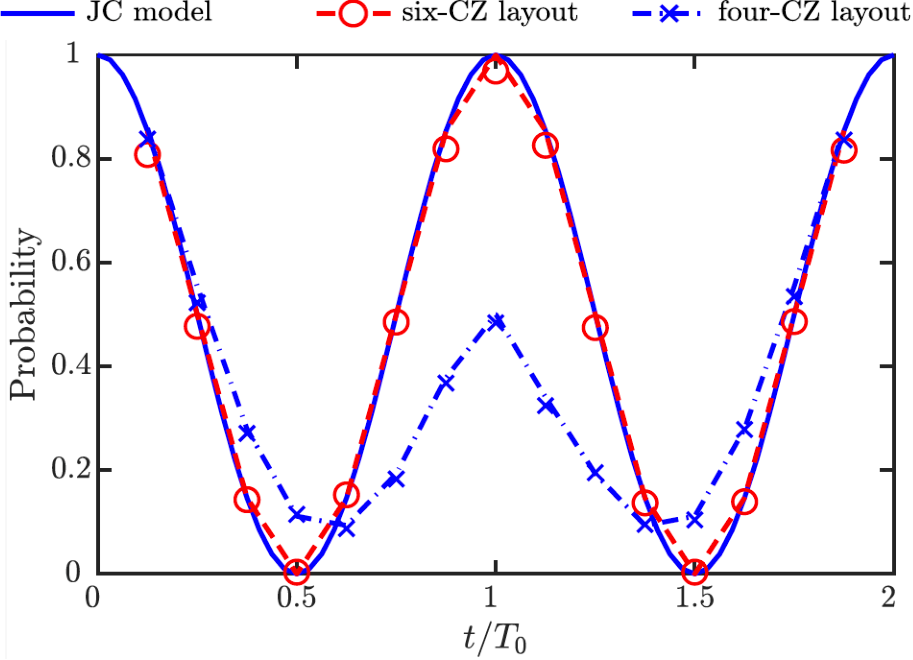}
    \caption{Probability of $\ket{0,\uparrow_{\tau}}$ vs $t/T_0$ from the synthesized unitary approach. Red dashed line (circles), six-CZ-gate layout from noiseless (noisy) qubits; and blue dash-dotted line (crosses), best four-CZ-gate layout from noiseless (noisy) qubits. The blue solid line is the analytical results for the JC model. All parameters are the same as in Fig.~\ref{fig8}. }
    \label{fig9}
\end{figure}
Using the optimized parameters for the four-CZ-gate layouts and the six-CZ-gate layout, we simulate the vacuum Rabi oscillation studied in Sec.~\ref{sssec:4B1}. Figure~\ref{fig9} shows the probability of the state $\ket{0, \uparrow_{\tau}}$ as a function of $t/T_{0}$ for these layouts. As can be seen, the six-CZ-gate layout generally outperforms the four-CZ-gate layouts, producing higher accuracy with respect to the ideal cavity case. An exception occurs near the beginning and end of the evolution, where the four-CZ-gate layout achieves comparable accuracy and even outperforms the six-CZ-gate layout on IBMQ, owing to its shallower circuit depth. Compared with the Suzuki-Trotter decomposition, the six-CZ-gate layout yields significantly improved results. In the Trotter approach, the case with $K_T = 1$ Trotter step (with six CZ gates) gives an average probability difference of $\Delta P_{\mathrm{tot}} = 0.30$, and the case with $K_T = 4$ (with 18 CZ gates) gives $\Delta P_{\mathrm{tot}} = 0.06$. For the six-CZ-gate layout, the average probability difference is reduced to $\Delta P_{\mathrm{tot}} = 0.019$. The device calibration data for the qubits used in these simulations are given in Table~\ref{tab4} of Appendix~\ref{sec:appa}. These results show that for quantum simulation of small-sized systems, the synthesized unitary approach can outperform Trotter decomposition at the cost of additional classical computational resources for circuit optimization.

\subsubsection{Diagnostic Benchmark of CZ-Gate Errors \label{sssec:4B3}}
\begin{figure}
    \centering
    \includegraphics[width=\columnwidth, clip]{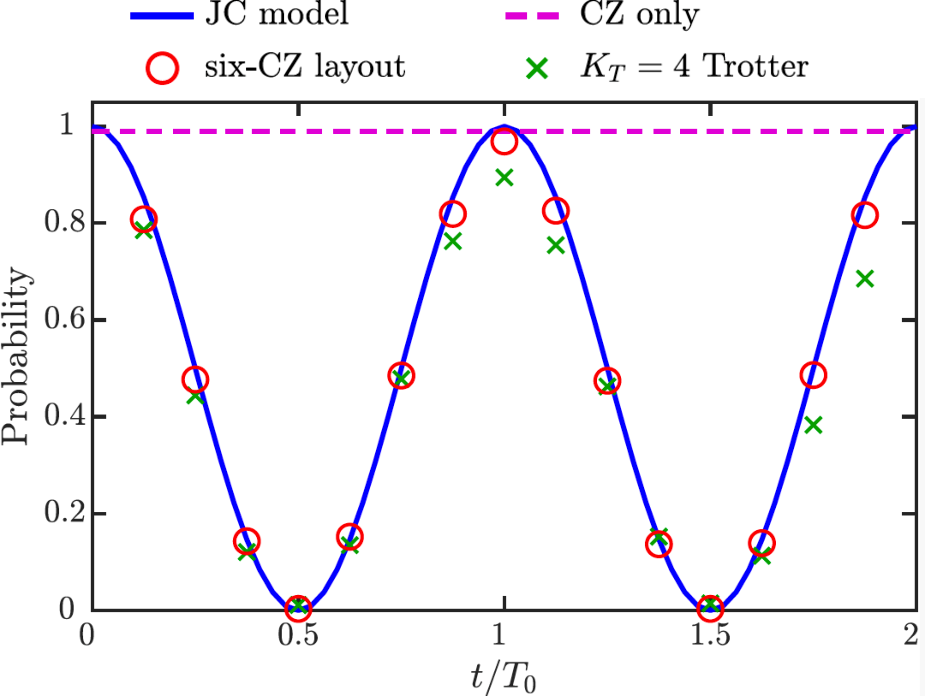}
    \caption{Probability of $\ket{0,\uparrow_{\tau}}$ vs $t/T_0$ for the six-CZ-gate synthesized unitary approach (red circles), the $K_T = 4$ Trotter results (green crosses), and the CZ-gate-only circuit (magenta dashed horizontal line). The blue solid line is the analytical results for the JC model. All parameters are the same as in Fig.~\ref{fig8}.}
    \label{fig10}
\end{figure}
On IBMQ hardware, two-qubit gate errors and readout errors constitute major sources of hardware noise with typical error rates $O(10^{-2})$ and $O(10^{-3})$, respectively,   whereas single-qubit gate errors are typically on the order of $10^{-4}$. Moreover, entangling gates generally require substantially longer execution times than single-qubit operations, increasing the accumulated effects of decoherence associated with finite relaxation ($T_1$) and dephasing ($T_2$) times. These considerations motivate us to examine the contribution of the two-qubit gates to the hardware error in the digital simulations.  For this purpose, we consider a diagnostic benchmarking circuit by removing all single-qubit rotations from the circuit in Fig.~\ref{fig7} and retaining only the two-qubit gates. On IBM Kingston, the native two-qubit gate is the CZ gate. Since $R_{\rm CZ}^2=I$, an even number of consecutive CZ gates ideally produces the identity operation. Consequently, deviations of the measured state probability from its initial value provide a direct diagnostic of the accumulated hardware errors associated with the CZ-gate sequence. We note that noise-estimation circuits based on gate sequences that reduce to the identity for noiseless qubits have been developed to characterize and mitigate two-qubit gate noise~\cite{UrbanekPRL2021}. In contrast, the CZ-gate-only circuit employed here is used only as a diagnostic benchmark to isolate the accumulated effect of CZ-gate errors and compare its dependence on the number of two-qubit gates with that of the full JC simulation.

The initial state of the benchmarking circuit is $\ket{0,\uparrow_{\tau}}$, identical to that used in the JC model simulations. We first apply six CZ gates, matching the number of two-qubit gates used in a single Suzuki-Trotter step and the six-CZ-gate layout of the synthesized unitary approach. We execute the circuit for 2000 shots and estimate the probability of the state. Figure~\ref{fig10} compares the measured probabilities obtained from the Suzuki-Trotter expansion with $K_T=4$, the synthesized unitary approach, and the CZ-gate-only benchmarking circuit. We find that the deviation of the probability in the benchmarking circuit from the ideal value of unity is comparable to that of the synthesized unitary approach, but much smaller than that of the Trotter expansion. This is because the $K_T=4$ Trotter circuit includes 18 CZ gates, while the benchmarking circuit and the synthesized unitary approach both include only six CZ gates. This observation supports our conclusion that CZ-gate errors constitute a dominant source of hardware error in the JC simulations. The qubit calibration data are given in Table~\ref{tab4} of Appendix~\ref{sec:appa}.

We further perform simulations consisting of $K_T$ repetitions of the six-CZ-gate sequence, corresponding to a total of $6K_T$ CZ gates, using both noiseless simulation and IBMQ hardware. The probability difference $\Delta P_{\rm CZ}$ between the hardware simulation of the benchmarking circuit and the ideal result of unity is plotted as a function of $K_T$ in Fig.~\ref{fig8}(b) (magenta squares). It can be seen that $\Delta P_{\rm CZ}$ increases approximately linearly with $K_T$, exhibiting a similar increasing trend as $\Delta P_{\rm e}$ of the Trotter expansion (orange crosses). This behavior further supports our analysis that the accumulation of CZ-gate errors is primarily responsible for the growth of the hardware error with increasing number of Trotter steps.
\vskip 5mm

\subsection{Circuit Implementation of Two-Excitation Dynamics \label{ssec:2excitation}}
Next, we perform digital quantum simulation of the Rabi oscillation in the two-excitation subspace $\{\ket{2,\downarrow_\tau},\ket{1,\uparrow_\tau}\}$ using $N=3$ qubits to represent the cavity mode. The simulation consists of two steps: (1) preparation of the initial state $\ket{D_2^3,\downarrow_\tau}$, corresponding to the cavity Fock state $\ket{2}$ with the qubit in the spin-down state; and (2) simulation of the Rabi oscillation in the two-excitation subspace.

For step (1), we employ the synthesized unitary approach to prepare the initial state using the following quantum circuit: 
\\[8pt]
{\begin{adjustbox}{width=0.48\textwidth}
\begin{quantikz}
\lstick[3]{{\large Cavity} \\ {\large in JC}} & \gate{U} & \qw & \qw & \qw & \qw & \gate[4]{R_{\mathrm{CZ}}} & \gate{U} & \qw & \qw & \qw & \qw & \qw & \qw & \qw\\
& \gate{U} & \qw & \qw & \gate[3]{R_{\mathrm{CZ}}} & \gate{U} & \qw & \qw 
    & \gate[3]{R_{\mathrm{CZ}}} & \gate{U} & \qw & \qw &  \gate[3]{R_{\mathrm{CZ}}} & \gate{U} & \qw\\
& \gate{U}  & \gate[2]{R_{\mathrm{CZ}}} & \gate{U} & \qw & \qw & \qw & \qw & \qw & \qw & \gate[2]{R_{\mathrm{CZ}}} & \gate{U} & \qw & \qw & \qw\\
\lstick {{\large Qubit} \\ {\large in JC}}  & \gate{U}  & \qw & \gate{U} & \qw & \gate{U} & & \gate{U} & & \gate{U} & \qw & \gate{U} & \qw & \gate{U} & \qw 
\end{quantikz}
\end{adjustbox}}
\\[8pt]
This circuit contains six CZ gates and achieves a state-preparation fidelity greater than $1-10^{-8}$ for noiseless qubits after optimization of the single-qubit gate parameters. We characterize the fidelity of this process on IBM Kingston by repeating the state-preparation circuit 20000 times and measuring the number of excitations $n_{\rm ex}$ in the prepared state. We obtain $\langle n_{\rm ex}\rangle=1.977$, in good agreement with the expected value of $\langle n_{\rm ex}\rangle=2$ for the target state. For step (2), we employ the Suzuki-Trotter approach with $K_T=4$ Trotter steps for each evolution time $t$. As discussed in Sec.~\ref{sssec:4B1}, the quantum circuit contains $34$ CZ gates in total. The data of the qubits are summarized in Table~\ref{tab6} of Appendix~\ref{sec:appa}. 

\begin{figure}[t]
    \centering
    \includegraphics[width=\columnwidth, clip]{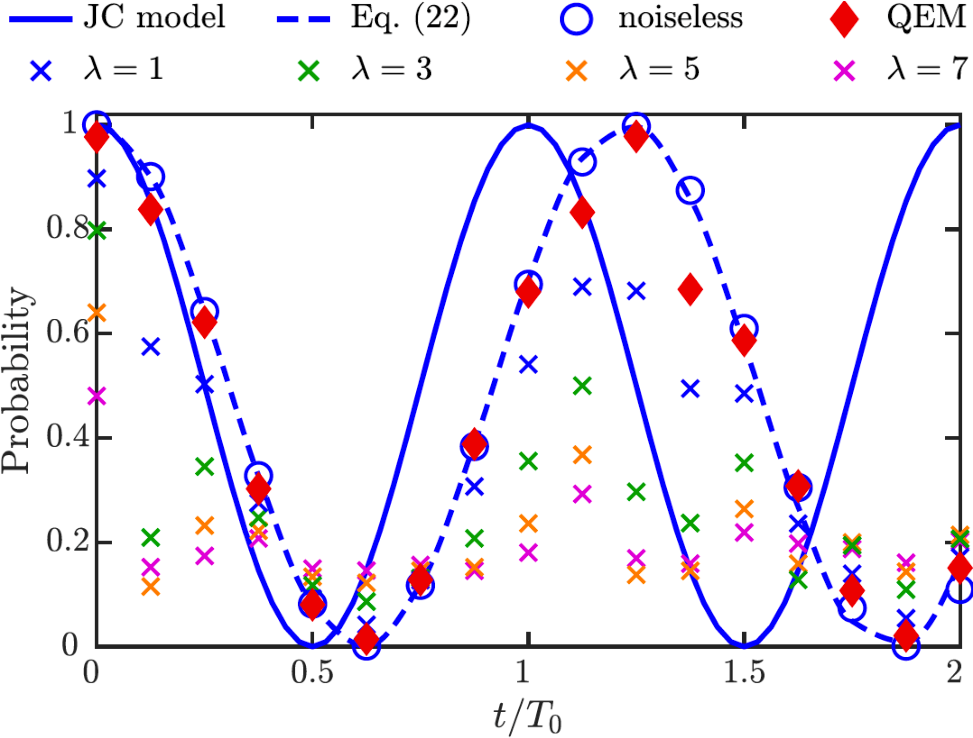}
    \caption{Probability of $\ket{2,\downarrow_\tau}$ vs $t/T_0$ for $N=3$ qubits.  Blue solid line: analytical results for the JC model; blue dashed line (circles): analytical [Eq.~\eqref{eq:PdownJC2N}] ($K_T=4$ Trotter) results for noiseless qubits; crosses: for noisy qubits with $\lambda=1,3,5,7$ (blue, green, orange, magenta) for ZNE; and red diamonds: error-mitigated results after M3 and ZNE. All other parameters are the same as in Fig.~\ref{fig8}.}
    \label{fig11}
\end{figure}
The results from the simulation are presented in Fig.~\ref{fig11}. As discussed in Sec.~\ref{ssec:JCMModel}, on resonance ($\omega_z=\omega_0$), the Rabi frequency in the two-excitation subspace of the JC model is $\Omega_{R2}=g_2=\sqrt{2}g$. Under the HP mapping with $N$ qubits, the corresponding Rabi frequency becomes $\Omega_{R2,c} = g_{N,2} = g\sqrt{2(N-1)/N}$. For $N=3$, this gives $\Omega_{R2,c}=\sqrt{4/3}g$, differing from the exact JC result. This difference is clearly observed in Fig.~\ref{fig11} and represents an intrinsic algorithmic error arising from the finite number of qubits in the HP mapping. The deviation of the hardware results from the noiseless qubit results (circles) is slightly larger than that observed in the single-excitation dynamics for $N=3$ in Fig.~\ref{fig8}(d). While the time-evolution circuits in the two simulations are identical, the two-excitation simulation requires an additional state-preparation circuit containing six CZ gates, resulting in greater accumulated hardware errors. After employing M3 and ZNE to mitigate the hardware errors, the error-mitigated results show substantially improved agreement with the noiseless qubit simulation. Because the JC Hamiltonian conserves the total excitation number, the dynamics studied here is confined to two-dimensional subspaces. The present approach can be extended to explore a larger part of the bosonic Hilbert space, for example, by considering the quantum Rabi model with counter-rotating terms or JC dynamics with coherent or squeezed initial states.

\section{Conclusion \label{sec:conclusion}}
In this work, we have demonstrated the digital quantum simulation of the HP transformation on a cloud-based superconducting quantum processor, where a bosonic mode is mapped onto and emulated by a qubit ensemble. We investigated representative single-mode bosonic dynamics, including the driven harmonic oscillator and bosonic squeezing, as well as the JC model describing light-matter interaction. For the driven harmonic oscillator, our results reveal the interplay between algorithmic errors arising from the finite size of the qubit ensemble and hardware errors dominated by measurement noise, leading to an optimal ensemble size for achieving high-fidelity simulations. The simulation of bosonic squeezing, which requires entangling gates under the HP transformation, provided a more involved test of the HP-based digital simulation. For the JC model, we employed both the Suzuki-Trotter decomposition and the synthesized unitary approach to simulate vacuum Rabi oscillations in the single- and two-excitation subspaces. The quantum hardware simulations exhibited increasing errors with circuit depth, and a diagnostic benchmark using CZ-only circuits supports our analysis that the accumulation of entangling-gate errors is a dominant source of hardware error. The simulation of the two-excitation Rabi oscillation directly illustrates the finite-$N$ algorithmic error of the HP mapping. We also showed that QEM techniques, including M3 and ZNE, substantially improve the hardware results. Overall, our study establishes a practical framework for implementing the HP transformation and simulating bosonic modes on digital quantum processors. The results demonstrate the distinct roles of finite-$N$ algorithmic errors and hardware-induced errors, as well as the effectiveness of QEM in improving the simulations. These methods can be extended to more complex spin-boson and multimode cavity models, paving the way for quantum simulation of many-body systems involving bosonic degrees of freedom in the NISQ era. Our findings further highlight the importance of algorithm-hardware co-design for optimizing digital quantum simulations involving bosonic modes.

\section*{Acknowledgments}
K.Y., A.M., and L.T. were supported by the National Science Foundation Awards No.~DMR-2037987 and No.~CICI-2530705. S.A. was supported by Japan's Ministry of Education, Culture, Sports, Science and Technology (MEXT) Quantum Leap Flagship Program under Grant No.~JPMXS0120319794, and the Center of Innovations for Sustainable Quantum AI
(Grant No. JPMJPF2221) from the Japan Science and Technology Agency (JST). K.Y. and L.T. thank NICT for hospitality.

\section*{Data Availability}
The codes and data used to generate the results presented in this work are publicly available at \cite{Yip2026data}. 

\appendix

\section{Circuit Data of IBMQ Backend \label{sec:appa}}
In this Appendix, we present device calibration data for the qubits of the IBM Kingston backend used in our quantum simulations. The backend calibration data at the time of execution are accessible through the IBMQ Platform~\cite{IBMQC}. The qubit coupling map can be obtained from the IBM backend configuration. The IBM Kingston processor employs fixed-frequency transmon qubits designed for stable operation. The typical operating frequencies of these qubits lie in the $4.5$--$5.5$-GHz range, with typical anharmonicities on the order of $300$--$350$ MHz. These values are representative of the device architecture. However, transition frequencies and anharmonicities for individual qubits are not publicly provided in the IBM Kingston backend calibration data, and therefore are not included in the tables below.

\begin{table}[hbt]
    \centering
    \begin{tabular}{|c|c|c|c|c|}
        \hline 
        qubit index & readout error & $SX$ error & $T_1(\mu s)$ & $T_2(\mu s)$ \\
        \hline 
        50 & 0.00244 & 0.00018 & 227 & 299\\
        60 & 0.00269 & 0.00042 & 193 & 299\\
        23 & 0.00281 & 0.00016 & 223 & 299\\
        69 & 0.00305 & 0.00018 & 154 & 299\\
        40 & 0.0033 & 0.0005 & 256 & 299\\
        88 & 0.00354 & 0.00027 & 186 & 299\\
        21 & 0.00366 & 0.00015 & 453 & 299\\
        117 & 0.00366 & 0.0002 & 272 & 299\\
        109 & 0.00378 & 0.00023 & 286 & 299\\
        44 & 0.00403 & 0.00026 & 174 & 299\\
        126 & 0.00403 & 0.00024 & 244 & 299\\
        48 & 0.00427 & 0.00019 & 184 & 299\\
        91 & 0.00427 & 0.0003 & 282 & 299\\
        15 & 0.00439 & 0.00015 & 352 & 299\\
        71 & 0.00439 & 0.00077 & 311 & 299\\
        98 & 0.00452 & 0.00034 & 220 & 299\\
    \hline 
    \end{tabular}
    \caption{Calibration data for the qubits used in Sec.~\ref{ssec:dho_implement}. The device characterization was performed on August 10, 2026 6:46 PM PST.} 
    \label{tab1}
\end{table}
For the driven quantum harmonic oscillator simulation, we utilized up to $16$ qubits. The qubit indices, readout error rates, single-qubit gate error rates for the SX gate, and decoherence times ($T_1$ and $T_2$) are summarized in Table~\ref{tab1}. 

\begin{table}[hbt]
    \centering
    \begin{tabular}{|c|c|c|c|c|}
        \hline 
        qubit index & readout error & $SX$ error & $T_1(\mu s)$ & $T_2(\mu s)$ \\
        \hline 
        $q_0$: 6 & 0.0092 & 0.00029 & 214 & 99\\ 
        $q_1$: 7 & 0.0132 & 0.00037 & 211 & 63\\ 
        $q_2$: 8 & 0.0236 & 0.00026 & 294 & 54\\ 
        $q_3$: 17 & 0.0051 & 0.00031 & 316 & 99\\ 
        $q_4$: 27 & 0.0095 & 0.00019 & 280 & 61\\ 
    \hline 
    \end{tabular}
    \caption{Calibration data for the qubits used in Sec.~\ref{ssec:squeezing_implement}. The device characterization was performed on August 11, 2026 1:46 PM PST.} 
    \label{tab2}
\end{table}

\begin{table}[hbt]
    \centering
    \begin{tabular}{|c|c|}
        \hline 
        qubit indices & CZ gate error\\
        \hline 
        $(q_0, q_1)$: (6, 7) & 0.0097\\ 
        $(q_1, q_2)$: (7, 8) & 0.0085\\ 
        $(q_1, q_3)$: (7, 17) & 0.0067\\ 
        $(q_3, q_4)$: (17, 27) & 0.0015\\ 
    \hline 
    \end{tabular}
    \caption{Calibration data for the two-qubit gates used in Sec.~\ref{ssec:squeezing_implement}. The device characterization was performed on August 11, 2026 1:46 PM PST.} 
    \label{tab3}
\end{table}
For the simulation of the squeezing operator, we utilized five qubits. The qubit indices, readout error rates, single-qubit gate error rates for the SX gate, and decoherence times ($T_1$ and $T_2$) are summarized in Table~\ref{tab2}. The error rates of the two-qubit gates are given in Table~\ref{tab3}. 

For the JC model, we employed $N=2, 3$ qubits to represent the cavity mode and one additional qubit to couple to the cavity. The calibration data for $N=2$ in Sec.~\ref{ssec:1excitation} are provided in Table~\ref{tab4}. The qubit indices, readout error rates, CZ gate error rates, single-qubit gate error rates for the SX gate, and decoherence times ($T_1$ and $T_2$) are summarized in this table.
\begin{table}[hbt]
    \centering
    \begin{adjustbox}{width=0.48\textwidth}
    \begin{tabular}{|c|c|c|c|c|c|}
    \hline
         qubit index & readout error & $R_{\mathrm{CZ}}$ error & $SX$ error & $T_1(\mu s)$ & $T_2(\mu s)$\\
         \hline
         O: 9 & 0.0034 & 0.033 & 0.0002 & 137 & 41 \\
         O: 11 & 0.0046 & 0.013 & 0.0003 & 287 & 119 \\
         A: 10 & 0.0073 &       & 0.0003 & 241 & 24 \\
         \hline
    \end{tabular}
    \end{adjustbox}
    \caption{Calibration data for the qubits used in $N=2$ simulations in Sec.~\ref{ssec:1excitation}. ``O'' refers to the qubits used to simulate the cavity mode, and ``A'' refers to the qubit coupled to the cavity mode. The device characterization was performed on August 13, 2026.} 
    \label{tab4}
\end{table}
The simulation of the JC model involves both single-qubit and two-qubit quantum logic gates. For single-qubit operations, only the SX (i.e., $\sqrt{X}$) gate is physically executed, whereas the $R_{Z}$ gate is implemented virtually as a frame change and requires no physical gate execution. Two-qubit gates have substantially longer execution times and are therefore more susceptible to circuit noise. On IBM Kingston, all two-qubit operations are realized using the CZ gate in combination with single-qubit rotations. Table~\ref{tab5} summarizes the total number of gates and the runtimes for the Suzuki-Trotter decomposition up to $K_T = 10$ Trotter steps and the synthesized unitary approach with $N=2$ qubits used to simulate the cavity mode.
\begin{table}[hbt]
    \centering
    \begin{adjustbox}{width=0.48\textwidth}
    \begin{tabular}{|c|c|c|c|}
    \hline
         {Trotter or synthesized unitary} & number of CZ gates & number of SX gates &  runtime ($\mu s$) \\
         \hline
         1 Trotter step & 6 & 30 & 0.9\\
         2 Trotter steps & 10 & 46 & 1.4\\
         3 Trotter steps & 14 & 62 & 1.9\\
         4 Trotter steps & 18 & 78 & 2.4\\
         5 Trotter steps & 22 & 94 & 3.0\\
         6 Trotter steps & 26 & 110 & 3.5\\
         7 Trotter steps & 30 & 126 & 4.0\\
         8 Trotter steps & 34 & 142 & 4.6 \\
         9 Trotter steps & 38 & 158 & 5.1 \\
         10 Trotter steps & 42 & 174 & 5.6 \\
         synthesized: four-CZ & 4 & 22 & 0.6 \\
         synthesized: six-CZ & 6 & 30 & 0.9 \\
         \hline
    \end{tabular}
    \end{adjustbox}
    \caption{Gate counts and execution runtimes for the simulations in Secs.~\ref{sssec:4B1} and \ref{sssec:4B2}. Runtimes are estimated using calibrated CZ gate and SX gate times ($68$ and $32$ ns, respectively) for IBM Kingston.}
    \label{tab5}
\end{table}
The calibration data for $N=3$ in Sec.~\ref{sssec:4B1} and Sec.~\ref{ssec:2excitation} are provided in Table~\ref{tab6}. Here the total number of CZ gates for $K_T$ Trotter steps is $8 K_T +2$. 
\begin{table}[hbt]
    \centering
    \begin{adjustbox}{width=0.48\textwidth}
    \begin{tabular}{|c|c|c|c|c|c|}
    \hline
         qubit index & readout error & $R_{\mathrm{CZ}}$ error & $SX$ error & $T_1(\mu s)$ & $T_2(\mu s)$\\
         \hline
         O: 150 & 0.0106 & 0.021 & 0.0002 & 238 & 313 \\
         O: 152 & 0.0094 & 0.009 & 0.0003 & 250 & 105 \\
         O: 138 & 0.0120 & 0.014 & 0.0001 & 329 & 99 \\
         A: 151 & 0.0055 &       & 0.0001 & 321 & 260 \\
         \hline
    \end{tabular}
    \end{adjustbox}
    \caption{Calibration data for the qubits used in $N=3$ simulations in Secs. ~\ref{sssec:4B1} and \ref{ssec:2excitation}. ``O'' refers to the qubits used to simulate the cavity mode, and ``A'' refers to the qubit coupled to the cavity mode. The device characterization was performed on August 20, 2026.} 
    \label{tab6}
\end{table}

\section{Quantum Error Mitigation\label{sec:appb}}
QEM techniques are widely employed in digital quantum simulation to compensate for hardware-induced errors without requiring full quantum error correction. In IBMQ's quantum processor used in this work, the dominant sources of error are errors associated with the entangling gates and measurement (readout) errors. To mitigate these effects, we employ M3 to reduce readout errors and ZNE to suppress errors associated with the entangling gates.

\subsection{Matrix-Free Measurement Mitigation \label{ssec:M3}}
M3 is an efficient technique for mitigating measurement (readout) errors in quantum processors~\cite{NationPRXQuantum2021}. Using calibration data obtained from the quantum hardware, M3 reconstructs the probability distribution of the measured bit strings by correcting for state-assignment errors, thereby improving the accuracy of measured observables without modifying the quantum circuit. Unlike conventional measurement mitigation methods that require constructing and inverting the full measurement calibration matrix, whose computational cost grows exponentially with the number of qubits, M3 employs a matrix-free approach that operates only within the subspace of the measured bit strings. This significantly reduces both the computational and memory costs while maintaining high mitigation accuracy for large quantum systems.

In this work, M3 is applied to all the simulations in Secs.~\ref{sec:dho} and \ref{sec:JCM}. For the driven harmonic oscillator, where the quantum circuit consists solely of single-qubit gates with shallow circuit depths, readout error is the dominant source of hardware error, and M3 alone yields a substantial improvement in agreement with the noiseless simulation.

\subsection{Zero-Noise Extrapolation \label{ssec:ZNE}}
ZNE is a widely used quantum error mitigation technique for estimating the noiseless expectation values of quantum observables without requiring detailed knowledge of the underlying noise~\cite{TemmePRL2017, Giurgica-TironIEEE2020}. The central idea is to intentionally amplify the hardware noise while preserving the ideal unitary evolution, and then extrapolate the measured observables to the zero-noise limit.

In this work, ZNE is employed to mitigate the errors associated with the quantum logic gates in the squeezing and the Jaynes-Cummings simulations. Noise amplification is implemented using unitary folding, in which the original unitary operator of the quantum circuit $U$ is replaced by
\begin{equation}
    U(k) = U(U^\dagger U)^{k},
\end{equation}
where $k$ is the number of folding layers. Since $U^\dagger U = I$, $U(k) \equiv U $ for all $k$ and the unitary evolution is unchanged for noiseless qubits. For noisy qubits, the quantum logic gates in the added folding layers increase the accumulated hardware error. A folding level of $k$ corresponds to a noise-scaling factor of $\lambda = 2k+1$. In our simulations, we employ folding levels $k=0, 1, 2, 3$, corresponding to noise-scaling factors of $\lambda=1,3,5,$ and $7$, respectively. The expectation value of each observable is measured at these noise levels. After applying M3 to mitigate measurement-induced errors at each noise level, the resulting expectation values are fitted using the quadratic polynomial 
\begin{equation}
    E(\lambda)=c_0+c_1\lambda+c_2\lambda^2,
\end{equation}
to derive the fitting parameters $c_0$, $c_1$, and $c_2$. The parameter $c_0$ then provides an estimate of the noiseless expectation value obtained by extrapolating to the zero-noise limit of $\lambda=0$.

\section{Trotter Decomposition \label{sec:TrotterMethod}}
The Trotter decomposition provides an effective method for approximating time evolution when the Hamiltonian consists of noncommuting terms that cannot be implemented simultaneously.  By dividing the total evolution time $t$ into $K_T$ small intervals of duration $t/K_T$, the total time-evolution operator can be approximated as a product of exponentials of the individual Hamiltonian components~\cite{Nielsen_Chuang_2010}.  

The accuracy of this approximation systematically improves with higher-order decompositions. The standard (first-order) Trotter expansion, $e^{(A+B)t} \approx (e^{At/K_T} e^{Bt/K_T})^{K_T}$, introduces a leading-order error that scales as $O(t^2/K_T)$. We adopt the first-order Trotter expansion in the implementation of the squeezing operator in Sec.~\ref{ssec:squeezing_implement}. In contrast, the symmetrized (second-order) Suzuki-Trotter decomposition, $e^{(A+B)t} \approx (e^{At/2K_T} e^{Bt/K_T} e^{At/2K_T})^{K_T}$, reduces the leading-order error to $O(t^3/K_T^2)$, providing a more accurate approximation of the time evolution~\cite{Suzuki1990PhysLettA}. We use the second-order Suzuki-Trotter decomposition in the implementation of the JC model in Secs.~\ref{ssec:1excitation} and \ref{ssec:2excitation}. 

%

\end{document}